\documentclass[12pt]{article}

\usepackage[]{inputenc}
\usepackage[T1]{fontenc}
\usepackage{hyperref} 
\usepackage{slashed}
\usepackage{color}

\usepackage{fullpage}

\usepackage{amsmath}
\usepackage{amssymb}
\usepackage[toc,page]{appendix}

\usepackage{amsthm}
\usepackage{float}
\usepackage{wrapfig}
\usepackage{caption}
\usepackage{graphicx}
\usepackage{pgfplots}
\usepackage{subcaption}

\def \be {\begin{equation}}
\def \ee {\end{equation}}
\def \bea {\begin{eqnarray}}
\def \eea {\end{eqnarray}}

\newcommand\HR[1]{{\it \color{purple}\bf  [#1 - HR]}}

\title{Quantum critical collapse to an extremal black hole}

\author{Hugo Lange and Harvey S. Reall \\
{\footnotesize Department of Applied Mathematics and Theoretical Physics, University of Cambridge,} \\{\footnotesize
Wilberforce Road, Cambridge CB3 0WA, United Kingdom} \\ {\footnotesize hl634@cam.ac.uk \qquad hsr1000@cam.ac.uk}}

\date{November 2026}

\begin{document}

\maketitle

\begin{abstract}
It has been emphasized recently that some extremal black holes are ``critical’’ solutions, lying at the threshold between gravitational collapse and dispersion. We study the effect of quantum gravity on critical solutions for the collapse of a thin  shell of charged dust. We consider the Hamiltonian that generates the shell equation of motion with respect to the exterior Killing time and coincides with the ADM energy on-shell. Near criticality, this Hamiltonian is shown to be canonically equivalent to that of a Liouville theory with negative potential. After quantization, this model predicts that a collapsing shell has a non-trivial probability of undergoing a quantum mechanical bounce when its energy above extremality is comparable to the ``breakdown scale”. The bounce probability approaches unity at extremality. Hence this model predicts that the critical surface in the classical theory is thickened quantum mechanically into a ``critical layer’’. The probability density function for the black hole energy is suppressed below the breakdown scale in the same way as found recently by Turiaci and Wu. 
\end{abstract}

\section{Introduction}

In General Relativity coupled to physically reasonable matter, consider some large class of asymptotically flat initial data. When the data is evolved, some solutions will describe gravitational collapse to form a black hole. Other solutions might be dispersive, settling down to flat spacetime at late time. A solution lying at the threshold between these two types of behaviour is referred to as a critical solution. A famous example of this is the spherically symmetric Einstein-massless-scalar field model, where critical solutions are nakedly singular \cite{Choptuik:1992jv}. 

For the collapse of charged matter, Kehle and Unger have observed that some critical solutions describe the formation of extremal charged black holes \cite{Kehle:2024vyt}. The simplest example of this is the spherically symmetric collapse of a shell of charged dust. Depending on its energy, it can either collapse to a Reissner-Nordstr\"om (RN) black hole or bounce without forming a black hole \cite{Boulware:1973tlq,FarrugiaHajicek1979,Proszynski1983}. The critical solution describes a shell that collapses to form an extremal RN black hole. The presence of a thin shell implies that these solutions are not smooth, but Kehle and Unger constructed analogous smooth solutions describing the spherically symmetric collapse of massless charged Vlasov matter \cite{Kehle:2024vyt}. (With massive charged Vlasov matter there are also ``unstable star'' critical solutions \cite{East:2025nfb}.)

Kehle and Unger emphasize that this observation has a surprising corollary. Consider a critical solution describing spherically symmetric gravitational collapse to form an extremal black hole. Now perturb the initial data for this solution. Because of the criticality of the solution, there are (generic) perturbations which result in the perturbed spacetime being dispersive instead of describing a black hole. Hence the “black hole property” of the critical solution is unstable! 
This is not some subtlety relating to the teleological definition of a black hole: the black hole property of the critical solution is something that can be verified locally. This is because for the solutions just described, after a finite time, the solution is exactly extremal RN so the event horizon is a marginally outer trapped surface. 

In other work \cite{Kehle:2022uvc}, Kehle and Unger have constructed solutions which describe violations of the third law of black hole mechanics \cite{Bardeen:1973gs}. These solutions describe a non-extremal black hole which absorbs charged matter to evolve in finite time to become an extremal RN black hole. These appear similar to critical solutions but there is an important difference. A non-extremal black hole possesses trapped surfaces hence a third-law violating solution must possess trapped surfaces. However, a critical solution cannot possess a trapped surface \cite{Kehle:2024vyt}. (If it did then any small enough perturbation of the solution would preserve this property, and so the perturbed solution would always describe a black hole, contradicting criticality.) Hence a third-law violating spacetime cannot be critical.

Recent work of Turiaci and Wu \cite{Turiaci:2026gtl} has investigated the third law of black hole mechanics in {\it quantum} gravity. They used Jackiw-Teitelboim (JT) gravity \cite{Jackiw:1984je,Teitelboim:1983ux} to study a model in which charged matter is added to a near-extremal RN black hole in an attempt to make it extremal. They found that quantum gravity effects make it harder to achieve extremality. Specifically, given some energy distribution for the infalling matter they determined the probability distribution for the energy of the final black hole. The result is that quantum gravity effects suppress the probability density near extremality, causing it to vanish as $\sqrt{E-Q}$ very close to extremality, where $E$ is the energy and $Q>0$ the charge of the hole.\footnote{We use units such that Newton's constant is $G=1$.} These quantum effects are important for  $E-Q \lesssim E_\star$ where
\be
\label{Estardef}
 E_\star = \frac{\hbar^2}{32 \pi^2 Q^3}
\ee
$E_\star$ is (up to numerical factors) the {\it breakdown scale} below which the emission of a typical Hawking quantum would appear to drive the black hole beyond extremality \cite{Preskill:1991tb}.

Turiaci and Wu also investigated whether it is possible to produce a final black hole with energy in the range $E-Q \lesssim E_\star$ where quantum gravity is expected to be important. They showed that if the matter is a shell with wavefunction well-localized in time then the probability density for the final black hole energy is always peaked well outside this range. To achieve a non-trivial probability of the final black hole having energy in this range requires the initial state of the shell to be highly delocalized in time, i.e., very quantum.

The goal of this paper is to investigate analogous questions for critical solutions. As emphasized above, third-law violating solutions are never critical so it appears that the approach of  \cite{Turiaci:2026gtl} cannot be used to investigate the critical threshold. In particular, this approach does not capture the fragility of the black hole property of critical solutions. 

We will revisit an old-fashioned approach to quantization of a thin shell \cite{Hajicek:1992mx,Hajicek:1992pu}. We consider a shell with a flat interior and a RN exterior. The equation of motion of the shell can be written in a form that resembles an energy conservation equation where the ADM energy $E$ is expressed in terms of the shell's radius and velocity. One can look for a Hamiltonian which reproduces this expression, and then try to  quantize the model. In doing this one has to decide which notion of time is conjugate to the Hamiltonian. Ref \cite{Hajicek:1992mx} used the shell's proper time and \cite{Hajicek:1992pu} used the interior Minkowski time (both for an uncharged shell). However, since $E$ is the ADM energy, it seems more natural to regard the Hamiltonian as generating translations in the RN time coordinate $t$, which coincides with the proper time of an observer at infinity. However, this results in a Hamiltonian that is rather complicated (see Appendix \ref{app:ham}) and therefore difficult to quantize. 

We will make progress by focusing on a shell that is close to the critical threshold, with mass just slightly less than its charge. In this case, there is a region of spacetime just outside the shell where the geometry can be approximated by ${\rm AdS}_2 \times S^2$, both above and below the threshold. The resulting phase space suffers from a problem arising because, for a black hole solution, $t$ is a good time coordinate both outside the event horizon but also inside the inner (Cauchy) horizon of the black hole. Part of the phase space describes a shell moving in the latter region, which we regard as unphysical (for reasons explained below). However, we can perform a canonical transformation that maps the physical region of phase space onto $\mathbb{R}^2$ and makes the Hamiltonian very simple: in dimensionless coordinates it is
\be
\label{hKY}
 h = \frac{K^2}{2} - e^{-Y}
\ee
where $h$ is the energy above extremality, $Y$ is ``position'' and $K$ is ``momentum'', although these are not simply related to the original position and momentum of the shell. This is the Hamiltonian of a particle of unit mass moving in a Liouville (exponential) potential with a {\it negative} coefficient. In this model, we will see that the region $Y \rightarrow -\infty$ where the potential is unbounded below corresponds to a shell at infinity in ${\rm AdS}_2$ whereas the asymptotic region $Y \rightarrow \infty$, where the potential vanishes, corresponds to a shell that collapses to form a black hole as\footnote{
Recall that $t$ is the static RN time coordinate so $t \rightarrow \infty$ at the future event horizon.} $t \rightarrow \infty$ (or emerges from a white hole as $t \rightarrow -\infty$). 

The quantization of Liouville theory with a negative potential has been discussed previously \cite{Fulop:1995di,Kobayashi:1996np,Fredenhagen:2003ut}, where the approach has been to find self-adjoint extensions of the Hamiltonian, corresponding to certain boundary conditions at $Y=-\infty$ that ensure that time evolution is unitary. However, in our problem we do not expect time evolution to be unitary. This is because classical shell trajectories can reach the boundary of AdS$_2$ in finite time. These trajectories are those that start or end at $Y=-\infty$. Hence, in a quantum description, we must allow probability to leave or enter the system at $Y=-\infty$. Time evolution will not be unitary because this is an open system. We will show that the problem can be understood as a scattering problem in which we specify a source corresponding to an incoming shell at $Y=-\infty$, which determines a reflected wave, corresponding to an outgoing shell at $Y=-\infty$ (i.e. a shell that has bounced), and a transmitted wave, propagating like an outgoing free particle at $Y=\infty$, which corresponds to a shell collapsing to form a black hole as $t \rightarrow \infty$. 

Using this model we see that a collapsing shell with definite energy $E<Q$, corresponding to a negative eigenvalue of $h$, will be reflected by the Liouville potential and return to infinity, i.e., such a shell definitely bounces without forming a black hole. For $E>Q$, corresponding to a positive eigenvalue of $h$, the shell can now reach $Y=\infty$, i.e., a black hole can form, as predicted classically. But there is also a non-zero probability that the shell will be reflected by the Liouville potential and bounce without forming a black hole. The bounce probability is
\be
\label{Pbounce}
 P_{\rm bounce} = \exp \left(-\sqrt{\frac{E-Q}{E_\star}}\right)
\ee
This is non-negligible only for $E-Q \lesssim E_\star$ and tends to $1$ as $E \rightarrow Q$. So our model predicts that quantum gravity effects qualitatively change the classical prediction when the energy above extremality is below the breakdown scale. The sharp classical transition at $E=Q$ between collapse and bounce is smeared out quantum mechanically into a transition occurring over the (very small) energy range $Q \le E \lesssim Q+E_\star$. One could say that the critical surface is replaced by a critical layer.

States of definite energy are completely delocalized in time so we will also consider localized wavepackets built from superpositions of energy eigenstates. A wavepacket with support entirely on negative $E-Q$ leads to a bounce with probability $1$. If the wavepacket has some support on positive $E-Q$ we find that the probability density function for the black hole energy is suppressed near extremality in exactly the same way found by Turiaci and Wu \cite{Turiaci:2026gtl}, i.e., by a factor $\sqrt{E-Q}$ when $E-Q \lesssim E_\star$. For a Gaussian wavepacket centred on the extremal value $E=Q$, the probability of bouncing and not forming a black hole is close to $1$ if the wavepacket is highly localized in energy and therefore very delocalized in time. If the wavepacket is highly localized in time then the bounce probability is $1/2$ because such a wavepacket has (energy) width large compared to $E_\star$, with half of the support in $E<Q$. In this case, if a black hole forms then it typically has $E-Q \gg E_\star$ as in \cite{Turiaci:2026gtl}; if a bounce occurs then the typical time that it takes for the shell to emerge is very long, much longer than the age of the Universe for a shell of solar mass.

An obvious quantity of interest is the position of the shell. Classically, this is a complicated function of $(Y,K)$. If one tries to promote this to an operator in the quantum theory then one encounters non-polynomial functions of an operator that is symmetric but not self-adjoint, for the same reason that the momentum operator on a half-line is not self-adjoint. As in the latter case, we can deal with this issue by enlarging the Hilbert space. This enables us to construct a self-adjoint position operator in the larger space. From the perspective of the physical Hilbert space, this defines a Positive Operator Valued Measure (POVM) \cite{Nielsen:2012yss} for the shell position. Using this we can study how the probability density for the position evolves over time.

This paper is organized as follows. In Section \ref{sec:classical} we give a classical analysis of the system culminating in the Hamiltonian \eqref{hKY}. In section \ref{sec:quantum} we explain how to quantize the system and study how quantum effects modify the classical predictions. Section \ref{sec:povm} explains how to construct a POVM for the position of the shell. Finally in section \ref{sec:discuss} we discuss our results further. 

\section{Classical shell motion}

\label{sec:classical}

\subsection{Shell equations}

Consider a spherically symmetric thin charged dust shell with mass $m$, charge $Q$ and energy $E$. The spacetime inside the shell is flat and the spacetime outside the shell is Reissner-Nordstr\"om:
\be
 ds^2 = -f(r) dt^2 + f(r)^{-1} dr^2 + r^2 d\Omega^2 \qquad f(r) = 1-\frac{2E}{r} + \frac{Q^2}{r^2}
\ee
The equations of motion of the shell are \cite{Boulware:1973tlq,FarrugiaHajicek1979, Proszynski1983}
\be
 \dot{t} = f(r)^{-1} \left(\frac{E}{m} - \frac{Q^2+m^2}{2mr}\right)
\ee
and
\be
\label{req}
 \dot {r}^2 =\frac{1}{m^2}\left( E-\frac{Q^2-m^2}{2r}\right)^2 - 1
\ee
where a dot denotes a derivative w.r.t. the shell's proper time. The RHS of \eqref{req} vanishes at two values of $r$ corresponding to where the shell ``bounces''. If $E \ge Q$ then the bounce occurs inside a black hole. We will be interested in a shell with $Q>m$. If such a shell collapses from infinity with energy $E>m$ then it will form a black hole if $E \ge Q$ and bounce without forming a black hole if $E<Q$. 

\subsection{Near-extremal shells}

We will take $Q>0$ and assume the shell satisfies
\be
\label{etadef}
 m = Q(1-\eta)
\ee
with $0<\eta \ll 1$. When at rest near infinity, such a shell is superextremal ($E<Q$) but it can be made subextremal ($E>Q$) through its kinetic energy. We will focus on shell trajectories with
\be
 E = Q(1 + \eta^2 \epsilon)
\ee
for $\epsilon$ of order $1$. Define a dimensionless radial coordinate $q$ and time coordinate $T$ by
\be
\label{qTdef}
 r = Q(1+ \eta q) \qquad \qquad t = \frac{QT}{\eta}
\ee
For $q$ of order $1$ we have $f = \eta^2(q^2 - 2\epsilon) + O(\eta^3)$ and the metric outside the shell is
\be
 ds^2  = Q^2 \left[ -(q^2 - 2 \epsilon) dT^2 + \frac{dq^2}{(q^2 - 2 \epsilon)}+ d\Omega^2 + O(\eta)\right]
\ee
which is the metric of ${\rm AdS}_2 \times S^2$ plus $O(\eta)$ corrections. The radius of curvature of AdS$_2$ and $S^2$ factors is $Q$. For $\epsilon<0$ we have AdS in global coordinates, for $\epsilon=0$ we have Poincar\'e coordinates and for $\epsilon>0$ we have Rindler coordinates. 
When $\epsilon>0$ there are horizons at $q=q_\pm$ where
\be
\label{qpmdef}
 q_\pm = \pm \sqrt{2\epsilon}
\ee
$q_+$ corresponds to a black hole or white hole horizon and $q_-$ corresponds to an inner future or past Cauchy horizon. We aim to describe the motion of the shell {\it outside} the black/white hole horizon, so when $\epsilon >0$ we will focus on the region $q >q_+$. 

The equations of motion of the shell become
\be
 Q\dot{T} = \frac{q+1+O(\eta)}{q^2 - 2 \epsilon+ O(\eta)}
\ee
\be
 Q^2\dot{q}^2 = 2q+2\epsilon+1 + O(\eta).
\ee
The latter equation shows that a bounce occuring within the AdS$_2$ region does so at $q=q_b$ where 
\be
\label{q_bounce}
 q_b = -1/2-\epsilon+O(\eta)
\ee
Combining these equations we obtain
\be
 \left( \frac{dq}{dT} \right)^2 = \frac{\left(q^2 - 2 \epsilon\right)^2\left( 2q+2\epsilon+1  \right)}{\left(q+1 \right)^2} + O(\eta)
\ee
The denominator vanishes at $q=-1$ because $\dot{T}$ vanishes there. This can only happen inside a black hole. If $\epsilon<1/2$ then we have $q_b>-1$ so the shell bounces before it reaches $q=-1$. 

We now want to find a Hamiltonian description of this system. The goal is to find a Hamiltonian $H(r,p_r)$ which generates time translations w.r.t. $t$ and takes the value $E$ when evaluated on a solution. Here $p_r$ is the momentum conjugate to $r$, which we need to determine. To relate to our dimensionless quantities we write
\be
\label{Hh}
 H(r,p_r) = Q(1+ \eta^2 h(q,p))
\ee
where $p = p_r/Q$ and $h(q,p)$ generates time translations w.r.t. $T$, and takes the value $\epsilon$ when evaluated on a solution. Since $dq/dT = \partial h/\partial p$, neglecting higher order corrections, to determine $h$ we have to solve
\be
 \left( \frac{\partial h}{\partial p} \right)^2 =  \frac{\left(q^2 - 2 h\right)^2\left( 2q+2h+1  \right)}{\left(q+1 \right)^2}
\ee
Assuming $q^2-2h>0$ (a consequence of $q>q_+$ for $h=\epsilon>0$), this equation can be integrated to give\footnote{The ``constant of integration'' corresponds to the freedom to shift $p$ by an arbitrary function of $q$, which is a canonical transformation, i.e., preserves $dq \wedge dp$, so there is no loss of generality in setting it to zero.}
\be
\label{htanh}
 h = -q-\frac{1}{2}+\frac{1}{2}(q+1)^2 \tanh^2 p
\ee
This Hamiltonian is (a) unbounded below as $q \rightarrow \infty$ and $p \rightarrow 0$ with $p = o(1/q)$; (b) not a convex function of $p$ so globally it is not the Legendre transform of any Lagrangian. The unboundedness corresponds to shells with $E<Q$ that bounce ``near infinity'' in AdS$_2$.

The classical equations of motion arising from this Hamiltonian are
\be
 \frac{dq}{dT} = (q+1)^2 \tanh p \, {\rm sech}^2 p \qquad \qquad \frac{dp}{dT} = 1 - (q+1) \tanh^2 p
\ee
The shell velocity $dq/dT$ vanishes at finite $q$ if either $p=0$ or $|p| =\infty$. The former case corresponds to a bounce. In the latter case, reaching $|p|=\infty$ requires infinite time $T$. This behaviour corresponds to a shell that collapses into a black hole (if $p<0$) or emerges from a white hole (if $p>0$). (Recall that $T$ is proportional to the RN time coordinate $t$, and a black/white hole horizon has $t \rightarrow \pm \infty$.)

\begin{figure}[htbp]
    \centering
    \includegraphics[width=0.5\textwidth]{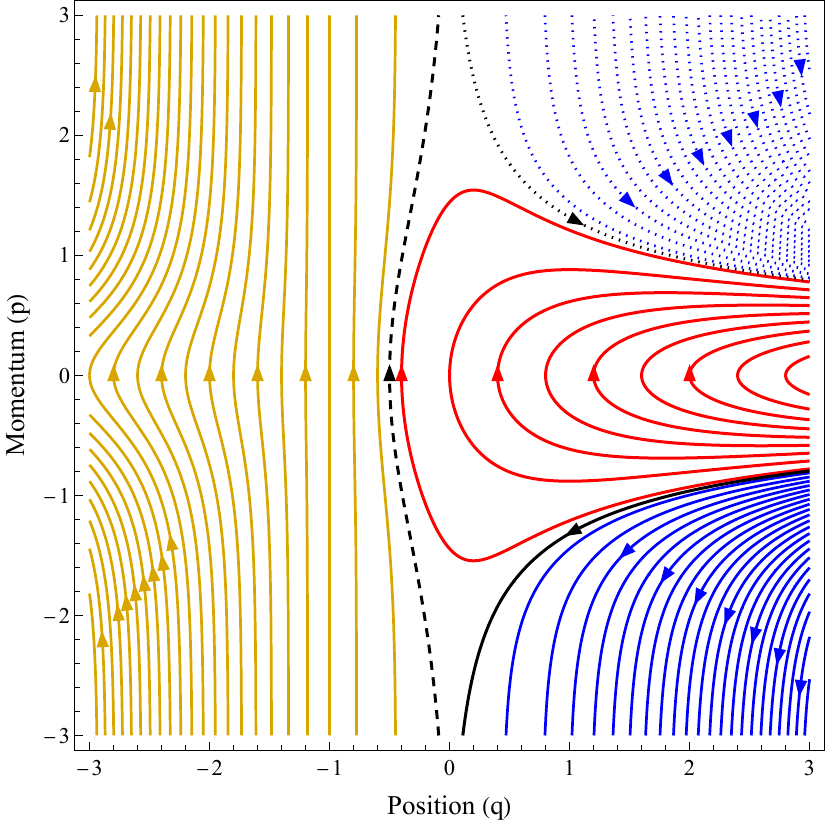}
    \caption{
    Contours of constant energy $h=\epsilon$ in $(q,p)$ coordinates. The arrows denote the direction of time evolution. The red curves have $\epsilon<0$ and describe shells incident from infinity that bounce and return to infinity. Curves with more negative $\epsilon$ are further to the right. Contours with $\epsilon>0$ are disconnected: for each $\epsilon$ there are $3$ curves shown in solid/dotted blue and in yellow. At the top right the dotted blue curves describe a white hole that emits a shell that expands to infinity. At the bottom right the solid blue curves describe a shell collapsing from infinity to form a black hole. On the left, the yellow curves describe a shell moving in the unphysical region behind an inner horizon: the shell collapses from a past inner horizon, bounces and then expands through a future inner horizon. Curves with $\epsilon=0$ are shown in black. The solid black curve corresponds to a critical trajectory: a shell collapsing to form an extremal black hole.
    The physical region of phase space is the region to the right of the dashed black curve.} 
    \label{fig:phase_space}
\end{figure}
Contours of constant $h=\epsilon$ are shown in Fig. \ref{fig:phase_space}. A trajectory with $\epsilon<0$ (shown in red) describes a shell that falls from infinity in AdS$_2$, bounces, and returns to infinity. In general there are three types of trajectory with $\epsilon>0$. One of these (solid blue) describes a shell collapsing to form a black hole ($q \rightarrow q_+$ as $T \rightarrow \infty$). Another (dotted blue) describes a shell expanding out of a white hole ($q \rightarrow q_+$ as $T \rightarrow -\infty$). The third type of trajectory (yellow) has $q \rightarrow q_-$ as $T \rightarrow \pm \infty$. It describes a shell moving in the region inside the inner horizon of a black hole. The shell emerges from a past inner horizon at $T=-\infty$, bounces (where $p=0$), and falls across a future inner horizon at $T=+\infty$. 

To see why this third type of $\epsilon>0$ trajectory arises, recall that our derivation of the Hamiltonian \eqref{htanh} assumed only that $q^2-2h>0$. When $h=\epsilon>0$ this assumption is satisfied both for $q>q_+$ and for $q<q_-$. We have not imposed any condition which selects the former region over the latter. Thus our phase space describes both regions. But these regions are disconnected from each other in physical space and our phase space does not describe shells moving in the region $q_-<q<q_+$ {\it between} the outer and inner horizons. For this reason we regard the region covered by the yellow curves in Fig. \ref{fig:phase_space} as unphysical. 
The physical region of phase space is the region to the right of the dashed black curve in Fig \ref{fig:phase_space}.

\subsection{Canonical transformation}

We will now perform a canonical transformation that achieves two goals. First, it is defined only on the physical region of phase space and maps this region to $\mathbb{R}^2$, eliminating the unphysical region. Second, it simplifies the Hamiltonian. After a sequence of intermediate steps described in Appendix \ref{app:canonical}, we were led to define a new coordinate $Y$ and momentum $K$ by
\be
\label{canonical}
 Y = -\log B \qquad K = -B\sinh p
\ee
where
\be
\label{Bdef}
 B = \frac{1}{\cosh p}\left(q+\frac{1}{1+\cosh p}\right)
\ee
This satisfies $dY \wedge dK = dq \wedge dp$ so the transformation is canonical. The map $(q,p) \rightarrow (Y,K)$ is defined only on the physical region of phase space, which is the region $B>0$. On this region it is smooth and invertible, with a smooth inverse
\be
\label{canonical_inv}
 p = -\sinh^{-1} \left(Ke^Y\right) \qquad q = e^{-Y}\cosh p - \frac{1}{1+\cosh p}
\ee
In the new variables the Hamiltonian takes the simple form \eqref{hKY}. It describes a particle of unit mass moving in an exponential (Liouville) potential with negative coefficient. The potential is unbounded below as $Y \rightarrow -\infty$, which must correspond to the region of phase space describing shells that bounce near infinity as discussed above. 

\begin{figure}[htbp]
    \centering
    \includegraphics[width=0.5\textwidth]{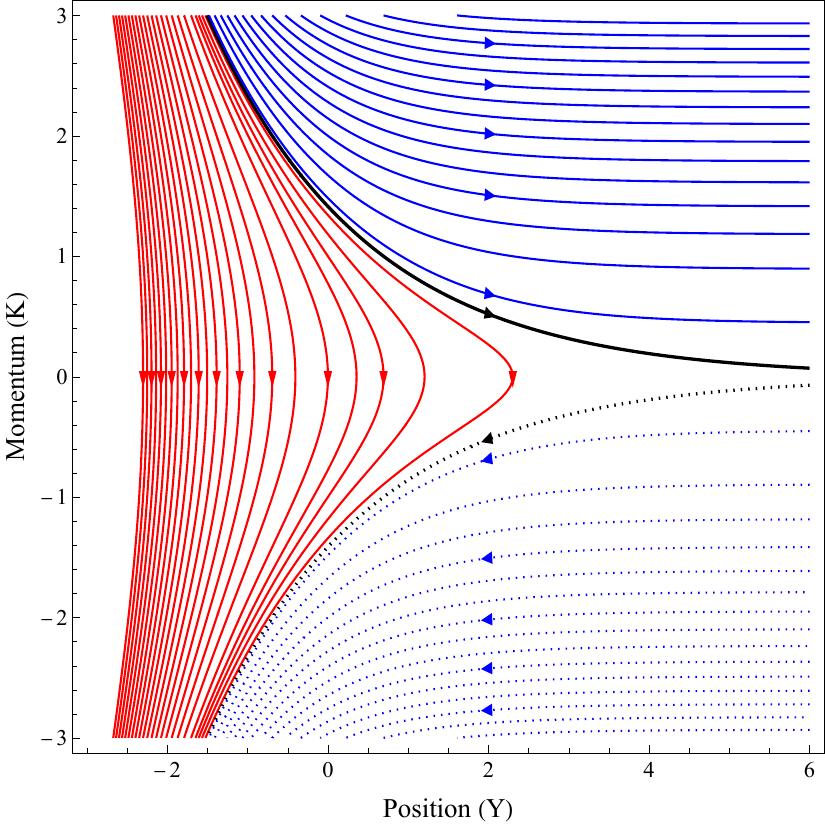}
    \caption{Contours of constant energy $h=\epsilon$ in $(Y,K)$ coordinates. The colour coding is the same as in Fig. \ref{fig:phase_space}. The solid blue lines correspond to black hole trajectories and the dotted blue lines correspond to white hole trajectories. The red trajectories correspond to a shell incident from infinity that bounces and returns to infinity. The black trajectories have $\epsilon=0$, the solid black line is the critical trajectory.}
    \label{fig:phase_space2}
\end{figure}

The classical trajectories in the $(Y,K)$ phase space are shown in \ref{fig:phase_space2}. In the new coordinates, a bounce ($p=0$, finite $q$) corresponds to $K=0$ at finite $Y$. A shell collapsing to a black hole corresponds to $K \rightarrow q_+$ and $Y \rightarrow \infty$. A shell emerging from a white hole corresponds to $K \rightarrow -q_+$ and $Y \rightarrow \infty$. A shell coming in from infinity corresponds to $K \rightarrow \infty, Y \rightarrow -\infty$ and a shell going out to infinity corresponds to $K \rightarrow -\infty, Y \rightarrow -\infty$. 

The equations of motion are straightforward to solve: up to the freedom to shift $T$ by a constant, any trajectory with $h =\epsilon$ is given by
\begin{subequations}
\begin{align}
 Y &=2\log \left[ \frac{1}{\sqrt{\epsilon}} \sinh \left( \sqrt{\frac{\epsilon}{2}} |T| \right) \right]  &&K = \sqrt{2\epsilon}\, {\rm coth}  \left( \sqrt{\frac{\epsilon}{2}} T \right)   &&\epsilon>0 \\
Y &= 2\log \left(\frac{|T|}{\sqrt{2}}\right)   &&K = \frac{2}{T}   &&\epsilon=0 \\
 Y &=2\log \left[ \frac{1}{\sqrt{-\epsilon}} \sin \left( \sqrt{\frac{-\epsilon}{2}} T \right) \right]   &&K = \sqrt{-2\epsilon}\, {\rm cot}  \left( \sqrt{\frac{-\epsilon}{2}} T \right)  &&\epsilon<0 \label{class_bounce}
\end{align}
\end{subequations}
For $\epsilon \ge 0$ the $T>0$ solutions describe a shell collapsing from infinity at $T=0$ to form a black hole and the $T<0$ solutions describe a shell expanding out of a white hole to reach infinity at $T=0$. For $\epsilon<0$ the range of $T$ is restricted to $0<T<\pi\sqrt{-2/\epsilon}$ with the shell collapsing from infinity at $T=0$, bouncing at $T=\pi/\sqrt{-2\epsilon}$, and returning to infinity at $T=\pi\sqrt{-2/\epsilon}$. 

\section{Quantization}

\label{sec:quantum}

\subsection{Introduction}

Write the Schr\"odinger picture state at time $t$ as $e^{-iQt/\hbar}|\Psi(T)\rangle$. In the dimensionless variables, the Schr\"odinger equation becomes
\be
 ig \frac{d}{dT} | \Psi(T) \rangle = h | \Psi(T) \rangle
\ee
where we have defined a dimensionless Planck constant
\be
\label{gdef}
 g = \frac{\hbar}{\eta Q^2} = \frac{\pi}{\eta S_0}
\ee
where $S_0 = \pi Q^2/\hbar$ is the Bekenstein-Hawking entropy of an extremal RN black hole of charge $Q$. For a black hole of astrophysical size, $S_0$ is enormous ($10^{76}$ for one solar mass) so if $\eta$ is only moderately small (say $10^{-6}$) then $g$ is tiny. 

Our expression for $h$ in terms of $(q,p)$ is complicated, so operator ordering issues will arise if we attempt to quantize it. Furthermore, the $(q,p)$ phase space suffers from the presence of the unphysical region. If we quantize in $(q,p)$ variables then the shell can quantum tunnel into the unphysical region. It is more natural to start from the expression for $h$ in terms of $(Y,K)$, where the Hamiltonian is simple and the unphysical region is absent. Therefore we define the quantum Hamiltonian by \eqref{hKY} where $Y,K$ are now operators satisfying the canonical commutation relation
\be
 [Y,K] = ig
\ee
Position space wavefunctions are denoted $\Psi(Y,T)$ and the Hilbert space will be taken as $L^2(\mathbb{R},dY)$. 

We do not want to quantize this system so that time evolution is unitary, for the reasons explained in the Introduction. So we do not follow the approach to quantization discussed in \cite{Fulop:1995di,Kobayashi:1996np,Fredenhagen:2003ut}. Instead we want to quantize as an open system.
We will see that if we prescribe a $T$-dependent ``source'' at $Y=-\infty$, corresponding to the wavefunction of an incoming shell at infinity, then the Schr\"odinger equation admits a well-posed initial-boundary value problem. Given the source, together with an initial state, then the equation determines a ``response'', corresponding to the wavefunction of an outgoing shell at infinity. 

In more detail, this works as follows. Wavefunctions of definite energy can be written in terms of Bessel functions \cite{Fulop:1995di,Kobayashi:1996np,Fredenhagen:2003ut} (see below) and an arbitrary superposition of these wavefunctions has the asymptotic behaviour
\be
\label{cddef}
 \Psi(Y,T) \approx c(T) H_0^{(1)}(z) + d(T) H_0^{(2)}(z) \qquad \qquad {\rm as}\,\,  Y \rightarrow -\infty
\ee
where $H_0^{(1,2)}$ are Hankel functions and
\be
\label{zdef}
 z = \frac{2\sqrt{2}}{g} e^{-Y/2}
\ee
A rigorous definition of $c(T),d(T)$ is given in Appendix \ref{app:cd}, where it is shown that $c(T)$ and $d(T)$ are well-defined for any $\Psi$ such that $\Psi,h\Psi \in L^2(\mathbb{R})$ with $\Psi$ twice differentiable in $Y$. We claim that $d(T)$ corresponds to the source and $c(T)$ to the response. To see this consider the probability current
\be
\label{Jdef}
 J(Y,T) = g {\rm Im }(\bar{\Psi}\partial_Y \Psi)
\ee
which appears in the conservation law
\be
\frac{\partial}{\partial T} |\Psi|^2 + \frac{\partial J}{\partial Y}=0
\ee
The global form of the conservation law is, defining $||\Psi(T)||^2 \equiv \int_{-\infty}^\infty dY \, |\Psi(Y,T)|^2$,
\be
\frac{d}{dT} ||\Psi(T)||^2 = -J(\infty,T)+J(-\infty,T)
\ee
At $Y=+\infty$ the potential vanishes so wavefunctions evolve as free particle wavefunctions. In particular the assumptions that $\Psi, h \Psi \in L^2(\mathbb{R})$ imply that $J(\infty,T)=0$ just as for a free particle. Probability only enters or leaves at $Y=-\infty$ where we have
\be
\label{Jlim}
 J(-\infty,T) = J_{\rm in}(T) - J_{\rm out}(T)
\ee 
where
\be
 J_{\rm in}(T) = \frac{g}{\pi} |d(T)|^2 \qquad \qquad J_{\rm out} = \frac{g}{\pi} |c(T)|^2 
\ee
From these expressions we see that $d(T)$ corresponds to a source of probability and $c(T)$ to a sink. Our problem is to solve the Schr\"odinger equation for prescribed $d(T)$ and a prescribed initial state, and to determine the resulting $c(T)$. This is well-posed as an initial-boundary value problem. To see this, use
\be
 \frac{d}{dT} ||\Psi(T)||^2 =  \frac{g}{\pi} \left( |d(T)|^2 - |c(T)|^2\right) \le \frac{g}{\pi}|d(T)|^2
\ee 
and hence
\be
||\Psi(T)||^2 \le ||\Psi(0)||^2 +\frac{g}{\pi}  \int_0^T  |d(T')|^2 dT'
\ee
Applying this inequality to the difference of two solutions with the same $d(T)$ and the same initial data $\Psi(Y,0)$ immediately gives uniqueness. Similarly, this inequality shows that the solution depends continuously on the initial data and on $d(T)$. Existence of the solution follows from a standard argument given in Appendix \ref{app:existence}.

Instead of specifying an initial wavefunction $\Psi(Y,0)$ at $T=0$ we will impose an initial condition at $T \rightarrow -\infty$
as follows. As noted above, for $Y \rightarrow \infty$ solutions will behave as free particle wavefunctions, describing a superposition of incoming and outgoing particles. By comparing with our classical phase space, we see that an outgoing particle at $Y =\infty$ corresponds to a shell approaching a black hole hole horizon (as $T \rightarrow \infty$) whereas an incoming particle at $Y=\infty$ corresponds to a shell emerging from a white hole horizon (as $T \rightarrow -\infty$). We are interested in what happens to a shell that collapses from infinity, with no white hole region present. Therefore we will impose the condition that the wavefunction at large $Y$ is ``purely outgoing''. The solution is then entirely determined by the source $d(T)$ associated with the shell collapsing from infinity. Viewed as a $1$-dimensional scattering problem, the source $d(T)$ corresponds to an incident wave, $c(T)$ to a reflected wave, and the outgoing particle to a transmitted wave.

\subsection{Energy eigenstates}

The time-independent Schr\"odinger equation is
\be
 -\frac{g^2}{2} \frac{d^2 \Psi}{dY^2} - e^{-Y} \Psi = \epsilon \Psi
\ee
For $\epsilon<0$ we require that the solution is normalizable. Such solutions exist for any $\epsilon<0$. They are not orthogonal in general. For $\epsilon>0$ we select the (non-normalizable) solution corresponding to an outgoing particle, i.e., $\Psi \sim \exp (i \sqrt{2\epsilon} Y/g)$ as $Y \rightarrow \infty$. For either sign of $\epsilon$ the resulting solution is
\be
 \Psi = A J_\nu (z)
\ee
where $A$ is a constant, $z$ is defined in \eqref{zdef} and 
\be
\label{nudef}
 \nu = -i \frac{\sqrt{8 \epsilon}}{g}
\ee
where we define the square root so that $\nu$ is positive for $\epsilon<0$. 

For $Y \rightarrow -\infty$ the large $z$ asymptotics of the Bessel function gives
\be
 \Psi \approx A \sqrt{\frac{g}{\pi \sqrt{2}}}\, e^{Y/4}\cos \left( \frac{2\sqrt{2}}{g} e^{-Y/2}+ \frac{i \pi}{g}\sqrt{2\epsilon}-\frac{\pi}{4} \right)
\ee
Notice that this is square integrable at $Y=-\infty$ although it oscillates increasingly rapidly as $Y \rightarrow -\infty$, which implies that $J_{\rm in}$ and $J_{\rm out}$ are non-zero. Inserting the time-dependence $e^{-i\epsilon T/g}$ and comparing with \eqref{cddef} gives
\be
\label{cdeigenstate}
 c(T) = \frac{A}{2} e^{-i\epsilon T/g}e^{-\pi \sqrt{2\epsilon}/g} \qquad \qquad d(T) = \frac{A}{2} e^{-i\epsilon T/g}e^{\pi \sqrt{2\epsilon}/g} 
\ee
and hence
\be
 c(T) = e^{-2\pi \sqrt{2\epsilon}/g} d(T)
\ee
which implies
\be
 J_{\rm out} = e^{-4 \pi {\rm Re}(\sqrt{2\epsilon})/g} J_{\rm in}
\ee
from which we infer that if we inject a shell of definite energy $\epsilon$ then it will bounce and return to infinity with probability
\be
 P_{\rm bounce}(\epsilon) = \begin{cases} 
 e^{-\sqrt{\epsilon/\epsilon_\star}} & {\rm if} \,\,\epsilon>0 \\
 1 & {\rm if }\,\, \epsilon \le 0
\end{cases}
\ee
where
\be
 \epsilon_\star = \frac{g^2}{32 \pi^2}
\ee
For $\epsilon<0$ this agrees with the classical picture: the shell definitely undergoes a bounce. We can also see this in the behaviour of the wavefunction: at large $Y$, i.e., small $z$, we have $\Psi\sim z^\nu \propto e^{-\nu Y/2}$ which decays exponentially for $\epsilon<0$. Hence the wavefunction vanishes in the large $Y$ region, i.e., the region that corresponds classically to a black (or white) hole. 

Now consider $\epsilon \ge 0$. For $\epsilon \gg \epsilon_\star$ the probability of a bounce is negligible. Again this agrees with the classical picture, which predicts that a shell with $\epsilon \ge 0$ will collapse to form a black hole and never bounce. However, if $\epsilon \sim \epsilon_\star$ then $P_{\rm bounce}$ is non-negligible and approaches $1$ as $\epsilon \rightarrow 0+$. This is a quantum effect that becomes important if we try to form a black hole with energy that is sufficiently close to extremality. This quantum bounce can be interpreted as quantum tunneling from a semiclassical ``shell collapsing to form a black hole'' solution into a semiclassical ``shell expanding from a white hole'' solution, even though, by construction, no white hole is actually present here.

This quantum effect become important for $\epsilon \lesssim \epsilon_\star$. Converting to dimensionful quantities we have
\be
 \frac{\epsilon}{\epsilon_\star} = \frac{E-Q}{E_\star}
\ee
where $E_\star$ is the breakdown scale defined in \eqref{Estardef}. Our expression for $P_{\rm bounce}$ now becomes \eqref{Pbounce} for $E \ge Q$. 

If we set $\epsilon=0$ then we have $P_{\rm bounce}=1$. So if we collapse a shell whose energy is exactly tuned to extremality then it will bounce rather than forming an extremal black hole. On the other hand, the classical prediction is that such a shell would definitely collapse to form an extremal black hole. So quantum effects dramatically modify the behaviour at the critical threshold. 

\subsection{Wavepackets}

Exact energy eigenstates are completely delocalized in time, as emphasized in \cite{Turiaci:2026gtl}. This makes them difficult to compare to classical trajectories corresponding to a shell entering from infinity at a definite time $T_0$. In order to facilitate such a comparison we need to consider wavepackets built from superpositions of energy eigenstates:
\be
 \Psi(Y,T) = \int_{-\infty}^\infty d\epsilon \,  A(\epsilon) e^{-i\epsilon T/g}J_{\nu}(z)
\ee
with $\nu$ defined by \eqref{nudef} and $z$ by \eqref{zdef}. It is convenient to define Fourier transforms of the response and source by
\be
 c(T) =  \int_{-\infty}^\infty d\epsilon \, e^{-i\epsilon T/g}\tilde{c}(\epsilon) \qquad \qquad  d(T) =  \int_{-\infty}^\infty d\epsilon \, e^{-i\epsilon T/g} \tilde{d}(\epsilon) 
\ee
We assume that $\tilde{d}(\epsilon)$ is smooth and decays rapidly as $\epsilon \rightarrow \pm \infty$. Repeated integration by parts then shows that $d(T)$ decays faster than any power of $T^{-1}$ as $|T| \rightarrow \infty$. 

From \eqref{cdeigenstate} we can read off
\be
 \tilde{c}(\epsilon) = \frac{A(\epsilon)}{2} e^{-\pi \sqrt{2\epsilon}/g} \qquad \qquad  \tilde{d}(\epsilon) = \frac{A(\epsilon)}{2} e^{\pi \sqrt{2\epsilon}/g}
\ee
hence the solution for the response in terms of the source is
\be
\tilde{c}(\epsilon)= e^{-2\pi \sqrt{2\epsilon}/g} \tilde{d}(\epsilon)
\ee
and the solution for the wavefunction in terms of the source is
\be
\label{psisol}
\Psi(Y,T)= 2\int_{-\infty}^\infty d\epsilon \,e^{-\pi \sqrt{2\epsilon}/g} \tilde{d}(\epsilon) e^{-i\epsilon T/g}  J_{\nu}(z)
\ee
To fix the normalization we will use the total probability coming in from infinity:
\be
 P_{\rm in} = \int_{-\infty}^\infty dT J_{\rm in}(T) = 2g^2 \int_{-\infty}^\infty d\epsilon \, |\tilde{d}(\epsilon) |^2
\ee
The total probability that leaves at infinity is similarly
\be
 P_{\rm out} = \int_{-\infty}^\infty dT J_{\rm out}(T) = 2g^2 \int_{-\infty}^\infty d\epsilon \, |\tilde{c}(\epsilon) |^2
\ee
An immediate consequence of these formulae is that if $\tilde{d}(\epsilon)$ is supported in $\epsilon \le 0$ then $|\tilde{c}(\epsilon)| =|\tilde{d}(\epsilon)|$ so $P_{\rm out} = P_{\rm in}$, i.e., if the shell is definitely super-extremal then it definitely bounces and returns to infinity. On the other hand, if $\tilde{d}(\epsilon)$ has any support in $\epsilon>0$ then we have $P_{\rm out}<P_{\rm in}$. Physically this is because the shell might collapse to form a black hole instead of bouncing and returning to infinity.  

We'll start by discussing the late time behaviour of the response. For large $T$ the integrand in $c(T)$ is highly oscillatory but there is no point of stationary phase. From our assumption that $\tilde{d}(\epsilon)$ is smooth with suitable decay at large $|\epsilon|$, the integral is dominated by the non-analytic behaviour at $\epsilon=0$ arising from the $\sqrt{\epsilon}$ dependence of the integrand. Expanding the integrand in powers of $\sqrt{\epsilon}$ gives, for $T \rightarrow \infty$
\be
\label{largeT_response}
 c(T) \approx 2\pi\sqrt{2\pi g} e^{i\pi/4}\tilde{d}(0)T^{-3/2}
\ee
so the late time response is governed by $\tilde{d}(0)$, i.e. the value of the source at the critical threshold. (The decay is faster if $\tilde{d}(0)=0$.) Using this, the fraction of the incoming probability that escapes after time $T$ is, for large $T$, 
\be
 P_{\rm in}^{-1} \int_T^\infty dT' J_{\rm out}(T') \approx \frac{2\pi^2}{T^2} \frac{|\tilde{d}(0)|^2}{\int_{-\infty}^\infty d\epsilon \, |\tilde{d}(\epsilon) |^2}
\ee
We can obtain similar results for the wave function \eqref{psisol}. At fixed $Y$, for large $T$ the integral is again dominated by the non-analytic behaviour at $\epsilon=0$ arising from the $\sqrt{\epsilon}$ dependence of the integrand (including through $\nu$). Expanding in powers of $\sqrt{\epsilon}$ gives, for $T \rightarrow \infty$ at fixed $Y$,
\be
\label{psi_late_time}
 \Psi(Y,T) \approx 2 \pi \sqrt{2\pi g}\, e^{i \pi/4}\, \tilde{d}(0) \, T^{-3/2} \, H_{0}^{(1)}(z)
\ee
so the probability density $|\Psi|^2$ decays as $T^{-3}$ at fixed $Y$. Some of this decay arises from the wavefunction dispersing to large $Y$, where it evolves as a free particle wavefunction. As noted above, an outgoing free particle at $Y=\infty$ corresponds to the shell approaching a black hole horizon. A free particle with velocity $V$ has energy $\epsilon = V^2/2$ so a black hole of energy $\epsilon$ corresponds at large $Y$ to a free particle moving with velocity $\sqrt{2\epsilon}$. So let's consider the behaviour of the wavefunction \eqref{psisol} at $Y = VT$ as $T \rightarrow \infty$. There is now a unique point of stationary phase at $\sqrt{2\epsilon} = V$ and the stationary phase approximation gives
\be
 \Psi(VT,T) \approx 2V \sqrt{\frac{2\pi g}{T}} \frac{e^{-\pi V/g}}{\Gamma(1-2iV/g)} \left( \frac{\sqrt{2}}{g}\right)^{-2iV/g}e^{i V^2 T/(2g)-i\pi/4}\tilde{d}(V^2/2)
\ee
so using $|\Gamma(1-ix )|^2=(\pi x)/\sinh(\pi x)$ for real $x$ gives
\be
 |\Psi(VT,T)|^2 \approx \frac{2g^2 V}{T}\left( 1 - e^{-4\pi V/g} \right) |\tilde{d}(V^2/2)|^2 \qquad \qquad {\rm as }\,\, T \rightarrow \infty
\ee
$|\Psi|^2$ is the probability density function for the coordinate $Y$. We can convert this into a probability density function for the energy $\epsilon = V^2/2$ of the final black hole using $|\Psi|^2 dY = T |\Psi|^2 dV = (T/V) |\Psi|^2 d\epsilon$. Normalizing by dividing by $P_{\rm in}$ we obtain the probability density function for the black hole energy:
\be
\label{fdef}
  f(\epsilon) = \frac{\left( 1 - e^{-\sqrt{\epsilon/\epsilon_\star}} \right) |\tilde{d}(\epsilon)|^2}{\int_{-\infty}^\infty d\epsilon' \, |\tilde{d}(\epsilon') |^2} \qquad \qquad \epsilon>0
\ee
From this we can obtain the probability density conditional on a black hole forming:
\be
 f_{\rm cond}(\epsilon) = \frac{f(\epsilon)}{\int_0^\infty d\epsilon' f(\epsilon')}  \qquad \qquad \epsilon>0
\ee
It is very interesting to compare this to the corresponding result of \cite{Turiaci:2026gtl} for collapse of a shell onto a near-extremal black hole. For $\epsilon \gg \epsilon_\star$, the energy of the black hole has the same probability distribution as the energy of the incoming shell. But for $\epsilon$ comparable to $\epsilon_\star$ the probability is suppressed by the prefactor $1 - e^{-\sqrt{\epsilon/\epsilon_\star}}$. In particular for $\epsilon \ll \epsilon_\star$ this prefactor is proportional to $\sqrt{\epsilon}$, exactly as in \cite{Turiaci:2026gtl}. In this approximation, $f_{\rm cond}(\epsilon)$ agrees exactly with the result of \cite{Turiaci:2026gtl}. We see that quantum effects cause the probability density for the black hole energy to vanish at extremality. Our model is very different from that of \cite{Turiaci:2026gtl} and it is reassuring that this prediction is common to both models. 

Reinstating dimensionful quantities, we can write $f(\epsilon)$ in the suggestive form
\be
 f(\epsilon) = 2e^{S_0}\sinh\left(\frac{1}{2}\sqrt{\frac{E-Q}{E_\star}}\right)\exp\left(-S_0 - \frac{1}{2}\sqrt{\frac{E-Q}{E_\star}}\right)\frac{|\tilde{d}(\epsilon)|^2}{\int_{-\infty}^\infty d\epsilon' \, |\tilde{d}(\epsilon') |^2}
\ee
where $S_0 = \pi Q^2/\hbar$ is the Bekenstein-Hawking entropy of an extremal black hole. The factor $2e^{S_0}\sinh\left(\frac{1}{2}\sqrt{\frac{E-Q}{E_\star}}\right)$ is the black hole density of states $\rho(E)$ predicted from including quantum gravity corrections associated with JT gravity \cite{Iliesiu:2020qvm}. To interpret the other factor, note that the Bekenstein-Hawking entropy of the black hole with energy $E$ can be written as
\be
S(E)=S_0 + \frac{1}{2}\sqrt{(E-Q)/E_\star}+O((E-Q)/(S_0 E_\star)). 
\ee
We can therefore write
\be
\label{f_sugg}
 f(\epsilon) = \rho(E)\overline{|{\cal M}|^2} \frac{|\tilde{d}(\epsilon)|^2}{\int_{-\infty}^\infty d\epsilon' \, |\tilde{d}(\epsilon') |^2}
\ee
with
\be
 \overline{|{\cal M}|^2} = e^{-S(E)}
\ee
The result \eqref{f_sugg} suggests that we can interpret $f(\epsilon)$ in terms of a scattering process, where a shell microstate of energy $\epsilon$ transitions to a black hole microstate. In this interpretation, ${\cal M}$ is the amplitude for the transition from the shell microstate to a given black hole microstate, the overbar denotes averaging over black hole microstates of the same energy, and $\rho(\epsilon)$ arises from summing over the black hole microstates with the same energy. This suggests that a typical shell to black hole microstate amplitude has size $|{\cal M}| \sim e^{-S/2}$, in agreement with a previous discussion \cite{Giddings:2009gj}. 

\subsection{Gaussian wavepacket}
\label{sec:gaussian}

To make the discussion more explicit, similarly to \cite{Turiaci:2026gtl}, we'll consider a Gaussian wavepacket built out of energy eigenstates centred at $\epsilon_0$, with width $\Delta$,
\be
 \tilde{d}(\epsilon) = d\,\exp\left(-\frac{(\epsilon-\epsilon_0)^2}{2\Delta^2} + \frac{i}{g}\epsilon T_0\right)
\ee
The corresponding $d(T)$ is a Gaussian centred at $T=T_0$ with width $\delta T \equiv g/\Delta$,
\be
 d(T) = d \sqrt{2\pi}\Delta\exp\left(-\frac{\Delta^2(T-T_0)^2}{2g^2} - \frac{i\epsilon_0 (T-T_0)}{g}\right)
\ee
For this wavepacket,
\be
\label{prob_dist_Gaussian}
 f(\epsilon) = \frac{1}{\Delta\sqrt{\pi}}\left(1-e^{-\sqrt{\epsilon/\epsilon_*}}\right)e^{-(\epsilon-\epsilon_0)^2/\Delta^2}
\ee

\subsubsection*{Wavepacket with $\epsilon_0=0$}

\begin{figure}[htbp]
    \centering
    \includegraphics[width=0.7\textwidth]{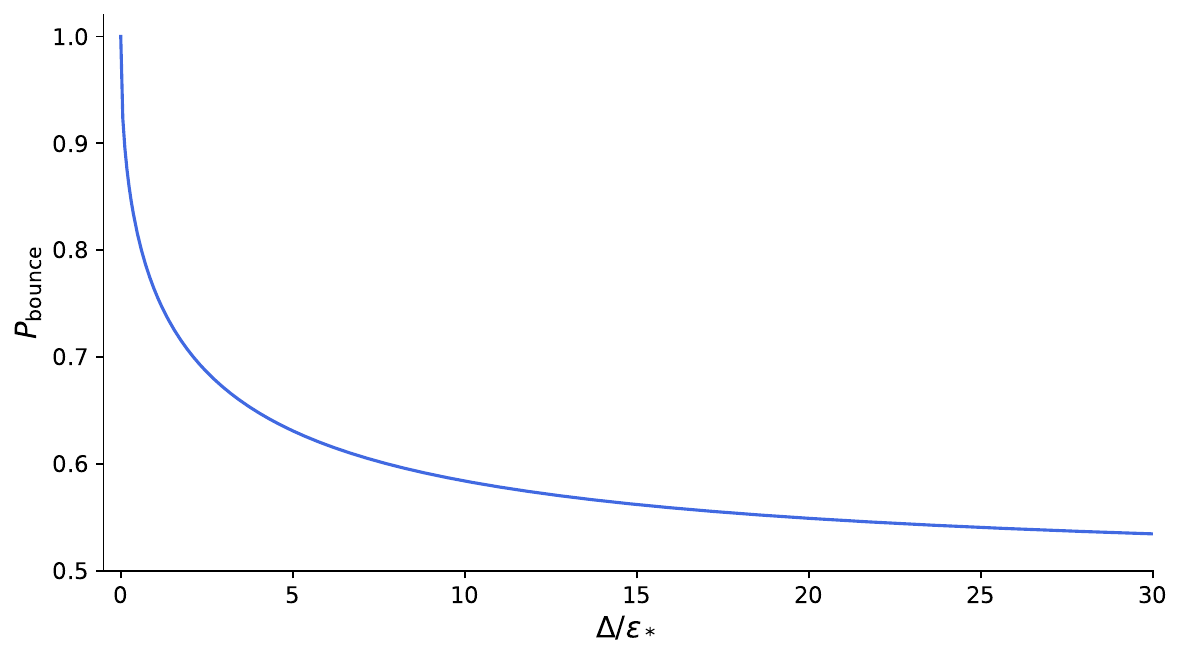}
    \caption{Plot of $P_\text{bounce}$ against $\Delta/\epsilon_*$ for $\epsilon_0=0$.}
    \label{fig:P_bounce}
\end{figure}

Our main interest is in the extremal threshold so let's first consider a wavepacket with $\epsilon_0=0$. Classically, a shell with $\epsilon=0$ collapses to form an extremal black hole. Quantum mechanically our shell bounces with probability
\be
 P_{\rm bounce} \equiv \frac{P_{\rm out}}{P_{\rm in}} 
 = \frac{1}{2} + \frac{1}{\sqrt{\pi}}\int_{0}^{\infty} d\xi \, \exp\left( -\xi^2 - \sqrt{\frac{\xi \Delta}{\epsilon_*}} \right)
\ee
where we have set $\xi=\epsilon/\Delta$. We plot $P_\text{bounce}$ as a function of $\Delta/\epsilon_*$ in Fig. \ref{fig:P_bounce}.
This probability is always at least $1/2$, because half of the Gaussian lies at negative $\epsilon$. If $\Delta/\epsilon_\star$ is small, i.e., the Gaussian is very narrow, then the bounce probability approaches $1$, in agreement with the behaviour of energy eigenstates. A very wide Gaussian has a bounce probability that approaches $1/2$. 

To compare with the classical picture, consider a wavepacket that is localized in time. The temporal width of the wavepacket in dimensionful units is $\delta t = (Q/\eta) \delta T = Q g/(\eta \Delta)$. We interpret ``localized in time'' to mean $\delta t/Q \lesssim 1$ so such a wavepacket has $\Delta \gtrsim \eta^{-1} g$ (i.e. $\delta T \lesssim \eta$). Equation \eqref{gdef} gives
\be
\label{localized}
  {\rm localized \,\, in \,\, time}\,\, \Rightarrow \frac{\Delta}{\epsilon_\star} \sim \frac{\Delta}{g^2} \gtrsim S_0 \gg 1 
\ee
where the final inequality assumes a macroscopic shell. From Fig. \ref{fig:P_bounce} it follows that the bounce probability is only slightly greater than $1/2$ for such a shell. Nearly all of this probability arises because half of the wavepacket is supported on negative $\epsilon$, i.e., on energy eigenstates describing bouncing shells. A typical shell has $|\epsilon| \lesssim \Delta$. For such a shell, with negative $\epsilon$, we can estimate\footnote{
We'll justify this estimate later in equation \eqref{tbounce2}.} how long it takes the shell to bounce using equation \eqref{class_bounce}, which gives a classical bounce time $T_{\rm bounce} \gtrsim 1/\sqrt{\Delta} = \sqrt{\delta T/g}$. In dimensionful units this gives
\be
\label{tbounce}
 \frac{t_{\rm bounce}}{Q} \gtrsim \sqrt{S_0} \sqrt{ \frac{\delta t}{Q}}
\ee
For a shell of solar mass, the huge factor $\sqrt{S_0}$ implies that if one sends in a wavepacket with $\delta t \lesssim Q$ and a bounce occurs then it typically takes an extremely long time for the shell to emerge, far longer than the age of the Universe.

\begin{figure}[htbp]
    \centering
    \includegraphics[width=0.8\textwidth]{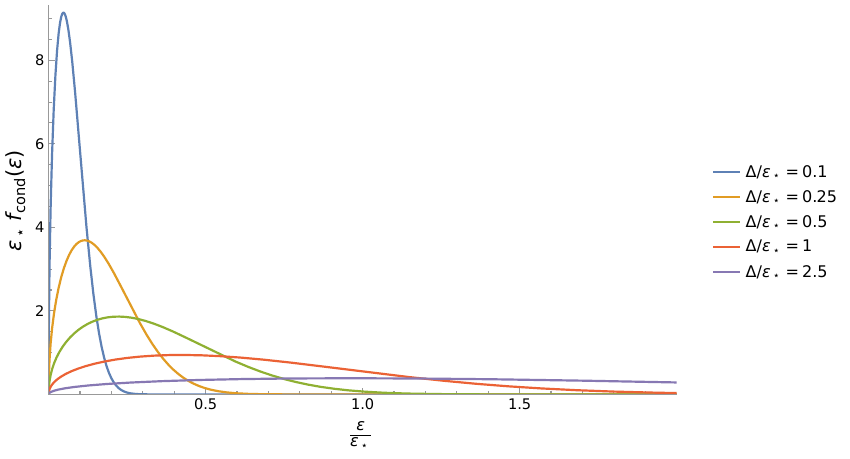}
    \caption{Plot of $f_{\rm cond}(\epsilon)$ for various values of $\Delta/\epsilon_*$ with $\epsilon_0=0$.}
    \label{fig:f_epsilon}
\end{figure}

If a black hole forms then its energy is distributed according to the function $f_\text{cond}(\epsilon)$. This is plotted for various values of $\Delta/\epsilon_*$ in Fig. \ref{fig:f_epsilon}. We've just seen that a shell that is localized in time has $\Delta/\epsilon_\star \gg 1$ and so for such a shell Fig. \ref{fig:f_epsilon} shows that the peak of the probability density function lies well above the breakdown scale. For the peak of the probability density to lie below the breakdown scale $\epsilon_\star$ requires that the width $\Delta$ is also below the breakdown scale. But this implies that $d(T)$ is highly delocalized in time, i.e., the initial state is highly quantum. These are very similar to the observations made in \cite{Turiaci:2026gtl}.

\subsubsection*{Wavepacket with $\epsilon_0 \le 0$}

Now consider arbitrary $\epsilon_0 \le 0$. We'll determine the form of the response function for the case of a source function that is highly localized in time, so \eqref{localized} holds. The response function is given by
\be
 c(T) = d\, \int_{-\infty}^{\infty} d\epsilon \, \exp\left(-\frac{(\epsilon-\epsilon_0)^2}{2\Delta^2}\right)\exp\left(-\frac{2\pi\sqrt{2\epsilon}}{g} - \frac{i\epsilon(T-T_0)}{g}\right)
\ee
We consider the positive and negative energy parts of this integral separately, calling these $c_+(T)$ and $c_-(T)$. The contribution from the negative energies is given by
\be
 c_-(T) = d\,\Delta\int_{-\infty}^{0}d\xi\, \exp\left(-\frac{1}{2}\left(\xi-\frac{\epsilon_0}{\Delta}\right)^2\right)\exp\left(\frac{i\sqrt{\Delta}}{g}\Phi(\xi)\right)
\ee
where we have substituted $\epsilon = \Delta \xi$, and where
\be
 \Phi(\xi) = -\sqrt{\Delta}\xi(T-T_0)-2\pi\sqrt{-2\xi}
\ee
We saw above that if the source is highly localized in time then $\Delta/g^2 \gg 1$, so we can use the stationary phase approximation. When $T>T_0$, $\Phi$ has a unique stationary point at $\xi_{\rm sp} = -2\pi^2/(\Delta(T-T_0)^2)$. The stationary phase approximation, for fixed $\sqrt{\Delta}(T-T_0)$, gives
\be
 c_-(T) \approx de^{i\pi/4}\,\exp\left(-\frac{1}{2}\left(\frac{2\pi^2}{\Delta(T-T_0)^2}+\frac{\epsilon_0}{\Delta}\right)^2\right)\exp\left(-\frac{2\pi^2i}{g(T-T_0)}\right)\frac{2\pi\sqrt{2\pi g}}{(T-T_0)^{3/2}}
\ee
We now consider $c_+(T)$. Since $\epsilon_0 \le 0$ we have
\be
 |c_+(T)| \le |d|e^{-\epsilon_0^2/(2\Delta^2)}\,\Delta \int_{0}^{\infty}d\xi \exp\left( -\frac{2\pi\sqrt{2\Delta\xi}}{g}\right) = \frac{|d|g^2}{4\pi^2}e^{-\epsilon_0^2/(2\Delta^2)}
\ee
which is therefore suppressed by a factor $O((g^2/\Delta)^{3/4})$ compared to $c_-(T)$. Hence 
\be
c(T) \approx c_-(T).
\ee
As $T \rightarrow \infty$ this reduces to \eqref{largeT_response}. We see that, in this approximation, the phase oscillates rapidly near $T=T_0$, while the frequency tends to zero as $T-T_0 \to \infty$. We can interpret this as being due to states in the superposition of the incoming pulse with large negative $\epsilon$ bouncing quickly back to infinity, while states with $\epsilon \lesssim 0$ take a long time to return to infinity. We can also see that the response peaks in amplitude at time
\be
 T \approx T_0 + \frac{2\pi}{\sqrt{-\epsilon_0 + \sqrt{\epsilon_0^2 + 3\Delta^2}}}
\ee
When $|\epsilon_0|/\Delta \gg 1$, we can expand in $\Delta/|\epsilon_0|$ to find that 
\be
 T \approx T_0 + \frac{2\pi}{\sqrt{-2\epsilon_0}}\left(1-\frac{3\Delta^2}{8\epsilon_0^2}\right)
\ee
To leading order, this matches the time taken for a classical bounce trajectory of energy $\epsilon_0 < 0$ to return to infinity after being initially injected at time $T_0$.

Meanwhile, for $\epsilon_0 = 0$, the peak is at
\be
\label{tbounce2}
 T \approx T_0 + \frac{2\pi}{3^{1/4}\sqrt{\Delta}}
\ee
which matches the time taken for a classical bounce trajectory with energy of order $\Delta$, justifying our use of the classical result in deriving \eqref{tbounce}. 

Finally we return to our assumption that $\sqrt{\Delta}(T-T_0)$ is held fixed in the stationary phase approximation. Converting to dimensionful quantities we have
\be
 \frac{t-t_0}{Q} \sim \sqrt{S_0} [\sqrt{\Delta}(T-T_0)] \sqrt{\frac{\delta t}{Q}}
\ee
and so our approximation is valid for time intervals of the order of \eqref{tbounce}. 

\subsubsection*{Wavepacket with $\epsilon_0 > 0$}

In this case, again assuming \eqref{localized}, if $\epsilon_0 \sim \Delta$ then the wavepacket has significant support on negative $\epsilon$ and the results for $c(T)$ are similar to the $\epsilon_0<0$ case. If $\epsilon_0 \gg \Delta$ then, for a wavepacket localized in time, \eqref{localized} implies that $\epsilon_0 \gg \Delta \gg  \epsilon_\star$, i.e., the wavepacket is supported on energies well above the breakdown scale so the probability of a bounce is negligible, hence the response function is very small. 

\section{Where is the shell?}

\label{sec:povm}

To answer this question classically we would use the canonical transformation \eqref{canonical_inv} from $(Y,K)$ coordinates back to $(q,p)$ coordinates. Quantum mechanically the question is more involved. In order to define $(q,p)$ in terms of $(Y,K)$ as self-adjoint operators we will need to enlarge our Hilbert space to a bigger space. From the perspective of the physical Hilbert space, this defines a Positive Operator Valued Measure (POVM) for measurements of $q$, as we will describe.

\subsection{Defining $q,p$ in the quantum theory}

We will break the classical transformation \eqref{canonical_inv} down into two steps. The first step is to define $B=e^{-Y}$ and $C = -Ke^Y$. The second step is to define $p = \sinh^{-1} C$ and then define $q$ using \eqref{Bdef}. In the quantum theory, we want operators corresponding to observables to be symmetric and so for the first step we define the operators
\be
 B = e^{-Y} \qquad \qquad C = -\frac{1}{2}(Ke^Y + e^Y K)
\ee
which satisfy $[B,C]=ig$. In terms of these operators, the Hamiltonian takes the form
\be
 h = \frac{1}{2}CB^2C-B-g^2/8
\ee
We now use $B$ as a new position coordinate, taking values in the half-line $(0,\infty)$. The map from wavefunctions in $L^2(\mathbb{R}, dY)$ to wavefunctions in $L^2((0,\infty),dB)$ is given by
\be
\label{transformation_Y_to_B}
 \chi(B) = B^{-1/2}\Psi(-\log B)
\ee
Acting on such a wavefunction we have $C = -ig \partial/\partial B$. It is well-known that the momentum operator is not self-adjoint on a half-line. Hence $C$ is not self-adjoint, and so it does not admit a spectral decomposition, and therefore the Born rule cannot be used to compute probabilities corresponding to measurements of $C$. We aim to write $(q,p)$ in terms of $(B,C)$ and so this problem will also occur for measurements of $(q,p)$. 

A natural (but non-unique) way to deal with this issue is to embed the physical Hilbert space $L^2((0,\infty),dB)$ in an enlarged Hilbert space $L^2(\mathbb{R},dB)$, by defining $\chi(B)= 0$ for $B<0$. The physical Hilbert space is the subspace of the enlarged space corresponding to wavefunctions satisfying the latter condition. In the enlarged Hilbert space, $C$ is self-adjoint, and so we can apply the Born rule to $C$ here. From the perspective of the physical Hilbert space, this procedure defines a positive operator valued measure (POVM), which we will discuss below.

For the second step of the transformation corresponding to \eqref{canonical_inv}, we define the operators
\be
\label{qp_quantum}
 q = \frac{1}{2}\{B, \sqrt{1+C^2}\} - \frac{1}{1+\sqrt{1+C^2}} \qquad \qquad p = \sinh^{-1}C
\ee
which satisfy $[q,p] = ig$. These operators are well-defined, and self-adjoint, in the enlarged Hilbert space. In this space the Hamiltonian is
\be
 h = \frac{1}{2}(q+1)(\tanh^2 p)(q+1)-q-\frac{1}{2}+\frac{g^2}{8}(1-\tanh^2 p)(5\tanh^2 p-1)
\ee
Comparing to the classical Hamiltonian \eqref{htanh}, our quantization procedure has specified a particular operator ordering and generated an $O(g^2)$ quantum correction.

The physical subspace of the enlarged Hilbert space is the subspace on which $B>0$. More formally, it is the positive spectral subspace of the operator $B$. The first equation of \eqref{qp_quantum} can be rearranged to write this operator in terms of the $(q,p)$ operators:
\be
 B= \frac{1}{2}\{ \text{sech}\, p, q \} + \frac{1}{\cosh p \, (1+\cosh p)}
\ee
In the enlarged Hilbert space we now switch from using wavefunctions $\chi(B)$ of the old position coordinate $B$ to wavefunctions $\tilde{\psi}(p)$ of the new momentum coordinate $p$. The map is given by
\be
\label{transformation_B_to_p}
 \tilde{\psi}(p) = \sqrt{\frac{\cosh p}{2\pi g}}\int_{-\infty}^\infty dB \exp\left[\frac{i}{g}\left(\tanh\frac{p}{2}-B\sinh p\right)\right]\chi(B)
\ee
Restricting this to wavefunctions in the physical Hilbert space, using \eqref{transformation_Y_to_B} (and $\chi=0$ for $B<0$), we can therefore write the map from wavefunctions in the physical space $L^2(\mathbb{R},dY)$ to wavefunctions in the enlarged space $L^2(\mathbb{R},dp)$ as
\be
\label{transformation_U}
 \tilde{\psi}(p) = \sqrt{\frac{\cosh p}{2\pi g}}\int_{-\infty}^{\infty} dY e^{-Y/2}\exp\left[\frac{i}{g}\left(\tanh\frac{p}{2}-e^{-Y}\sinh p\right)\right] \Psi(Y)
\ee
In the case of the energy eigenfunctions, $\psi_\epsilon(Y) = J_\nu(z)$, we can do this integral explicitly. This gives
\be
\label{eigenstates_p}
 \tilde{\psi}_\epsilon(p) = \sqrt{\frac{|\coth p|}{2}}\exp\left( \frac{i}{g}\coth p - \frac{\pi i}{4}(1+\nu)\text{sgn}(p) \right) J_{\nu/2}\left( \frac{1}{g|\sinh p|} \right)
\ee
Let's write $\mathcal{H}_\text{big} \equiv L^2(\mathbb{R},dp)$ for the space of momentum wavefunctions of the enlarged Hilbert space. The transformation \eqref{transformation_U} maps physical wavefunctions onto a subspace $\mathcal{H}_\text{phys} \subset \mathcal{H}_\text{big}$ corresponding to the positive spectral subspace of $B$. The projection onto this subspace, acting on momentum wavefunctions, can be determined explicitly as follows. Let $U:L^2(\mathbb{R},dY) \to \mathcal{H}_\text{big}$ be the operator enacting the transformation given in \eqref{transformation_U}, so the range of $U$  is $ \mathcal{H}_\text{phys}$. $U$ preserves norms, so is bounded, and satisfies $U^\dagger U = \mathbb{I}$ on $L^2(\mathbb{R},dY)$. The adjoint $U^\dagger$ is therefore also bounded and is defined everywhere on $\mathcal{H}_\text{big}$. Then $P = UU^{\dagger}$ is a projection operator mapping $\mathcal{H}_\text{big} \to \mathcal{H}_\text{phys}$, which acts on wavefunctions $\phi \in \mathcal{H}_\text{big}$ by
\be
\label{proj_op}
 \int_{-\infty}^{\infty}dp'\, P(p,p')\phi(p') = \frac{1}{2}\phi(p) - \frac{i}{2\pi}\text{PV}\int_{-\infty}^{\infty}dp'\, \frac{\sqrt{\cosh p \,\cosh p'}e^{\frac{i}{g}(\tanh (p/2) - \tanh (p'/2))}}{\sinh p - \sinh p'}\phi(p')
\ee
where $\text{PV}$ denotes the Cauchy principal value, taken at $p'=p$. $P$ is simply the momentum space representation of $\Theta(B)$ where $\Theta$ is the Heaviside step function.

\subsection{Position space wavefunction}

Equation \eqref{transformation_B_to_p} gives the momentum space wavefunction of a physical state in the enlarged Hilbert space. Its position space wavefunction is obtained by Fourier transforming in the usual way:
\be
 \psi(q) = \frac{1}{\sqrt{2\pi g}}\int_{-\infty}^\infty dp \, e^{ipq/g}\tilde\psi(p)
\ee
Let's now discuss the interpretation of this wavefunction. Recall that $q$ is not a self-adjoint operator in the physical Hilbert space, only in the enlarged space. In the enlarged space it is natural to use the Born rule to determine the probability of measuring the shell's position to lie in some interval, or union of intervals, $I \subset \mathbb{R}$ as
\be
 P(q \in I) = \int_I dq |\psi(q)|^2 = \langle \psi |R_I| \psi  \rangle 
\ee
where $R_I$ is the projection
\be
 R_I = \int_I dq |q \rangle \langle q|
\ee
$R$ does {\it not} preserve the physical subspace of the enlarged Hilbert space. However we can define 
\be
 F_I = P R_I P
\ee
where $P$ is the projection onto the physical subspace. This defines an operator on the physical Hilbert space and we have
\be
 P(q \in I) = \langle \psi| F_I | \psi \rangle
\ee
This construction defines a Positive Operator Valued Measure (POVM)  \cite{Nielsen:2012yss} for measurements of the shell's position. The operators $F_I$ are not projections in general. Therefore if we wish to talk about the state of the shell following a measurement of its position then we must supply a rule for how to update the state after a measurement. A simple rule is the square root prescription \cite{Nielsen:2012yss}. Consider a set of intervals $\{I_n \}$ that partition the real line and a measuring apparatus that can distinguish these intervals. If we measure $q$ and find that $q \in I_n$ then the shell's state should be updated to $\sqrt{F_{I_n}}|\psi \rangle$ and normalized. This generalizes the usual ``project onto the appropriate eigenspace'' rule for measurement of a self-adjoint operator.

In this construction, $|\psi(q)|^2$ is the probability density for a measurement of $q$, as in basic quantum mechanics. However, we need to be careful in interpreting a measurement of $q$ as telling us about the shell's position. Let's say we measure $q$, find $q \in I$ ($I=I_n$ for some $n$), and then immediately repeat the measurement. The above rules give the result of finding $q\in I$ again as
\be
\label{repeat}
 P_{\rm repeat}(q \in I) = \frac{\langle \psi |F_I^2 | \psi \rangle}{\langle \psi | F_I| \psi\rangle }
\ee
If this is close to $1$ then it increases our confidence in the result of the measurement. If it is not close to $1$ then we might ascribe the result of one or both measurements to a fluctuation or error in the measuring apparatus. We will return to this issue below. 

With these caveats in mind we will now present some plots showing the time evolution of $|\psi(q,T)|^2$, the probability density for the shell position. We will focus on a source function $d(T)$ described by a Gaussian wavepacket as described in section \ref{sec:gaussian}. 
Recall that $g$ is defined by \eqref{gdef}. For a macroscopic shell $g$ is incredibly tiny. However we will set $g=0.1$ to make generating plots tractable. For the wavepacket we will take $T_0=0$ and $\Delta=1$ so $\delta T = 0.1$. This corresponds to a source wavepacket that is injected from the AdS$_2$ boundary at a time centred on $T=0$.  Recall (above \eqref{localized}) that the condition for the wavepacket to be well-localized in time is $\delta T \lesssim \eta$ which requires $\eta \gtrsim 0.1$ here.

\begin{figure}[htbp]
    \centering
    \includegraphics[width=0.8\textwidth]{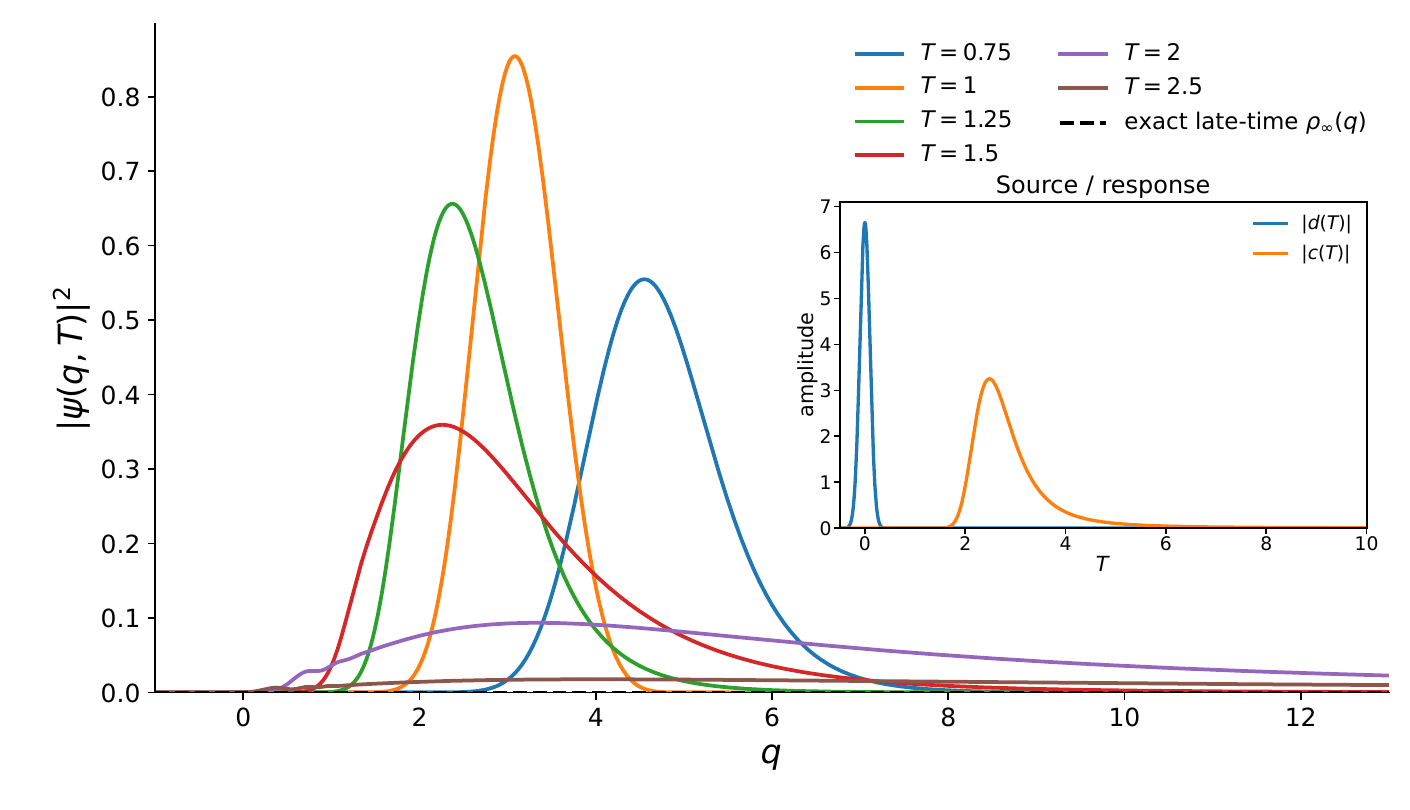}
    \caption{Evolution of the probability density arising from a source described by a Gaussian wavepacket with $T_0 = 0$, $\Delta = 1$, $g=0.1$, and $\epsilon_0 = -3$. This corresponds to a shell which bounces and returns to infinity. The inset shows the time-dependence of the source and response functions.
    \label{fig:early_time_gaussians_-3}
    }
\end{figure}
Fig. \ref{fig:early_time_gaussians_-3} shows results for $\epsilon_0=-3$, corresponding to a shell which bounces and returns to infinity with very high probability. Classically, the time for a shell with $\epsilon=-3$ to bounce and return to infinity is $\pi \sqrt{2/3} \approx 2.6$ from equation \eqref{class_bounce}. The location of the bounce is $q_b = 2.5$ from \eqref{q_bounce}. Quantum mechanically, the centre of the wavepacket evolves in rough agreement with these classical predictions. The width of the wavepacket is also seen to increase significantly at late time. This corresponds to lower energy shells taking longer to bounce than higher energy shells. 

\begin{figure}[htbp]
    \centering
    \includegraphics[width=0.8\textwidth]{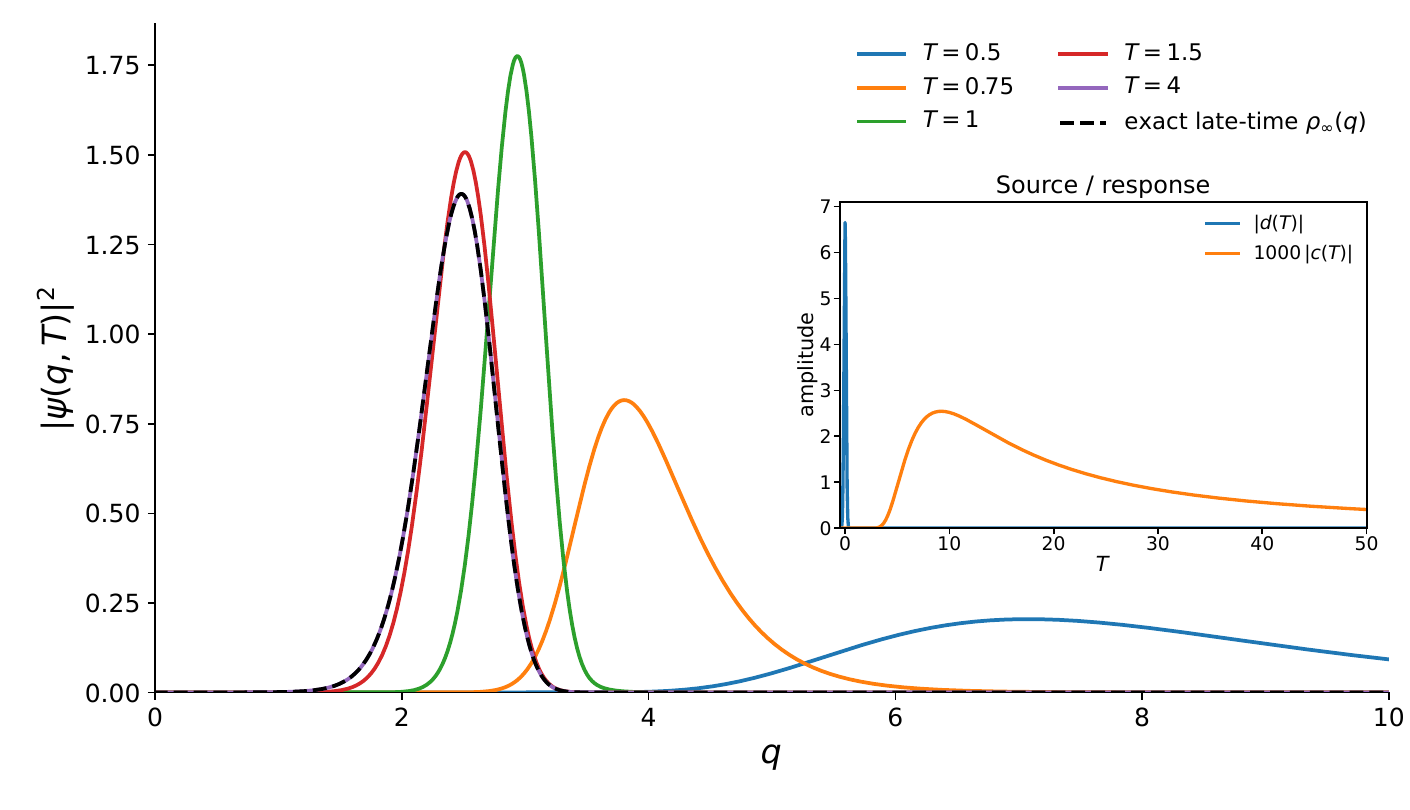}
    \caption{Similar to Fig. \ref{fig:early_time_gaussians_-3} with $T_0 = 0$, $\Delta = 1$, $g=0.1$, and $\epsilon_0 = 3$. This corresponds to a shell which collapses to form a black hole. In the inset, the response is multiplied by a factor of $1000$ so that it is visible on the plot.
    \label{fig:early_time_gaussians_3}
    }
\end{figure}
Fig. \ref{fig:early_time_gaussians_3} shows results for $\epsilon_0=+3$ ($\gg \epsilon_\star \sim 3 \times 10^{-5}$), corresponding to a shell that collapses to form a black hole with high probability. A classical shell with $\epsilon=3$ would form a black hole of radius $q_+ = \sqrt{6}=2.4$. Quantum mechanically we see that the wavefunction is indeed localized around $q\approx 2.4$ at late time. The dashed black curve is obtained from an analytical result for $T=\infty$ that we will derive in the next section. There is a tiny probability that the shell will bounce and return to infinity, this corresponds to the very small response shown in the inset of Fig. \ref{fig:early_time_gaussians_3}.

\begin{figure}[htbp]
    \centering
    \includegraphics[width=0.8\textwidth]{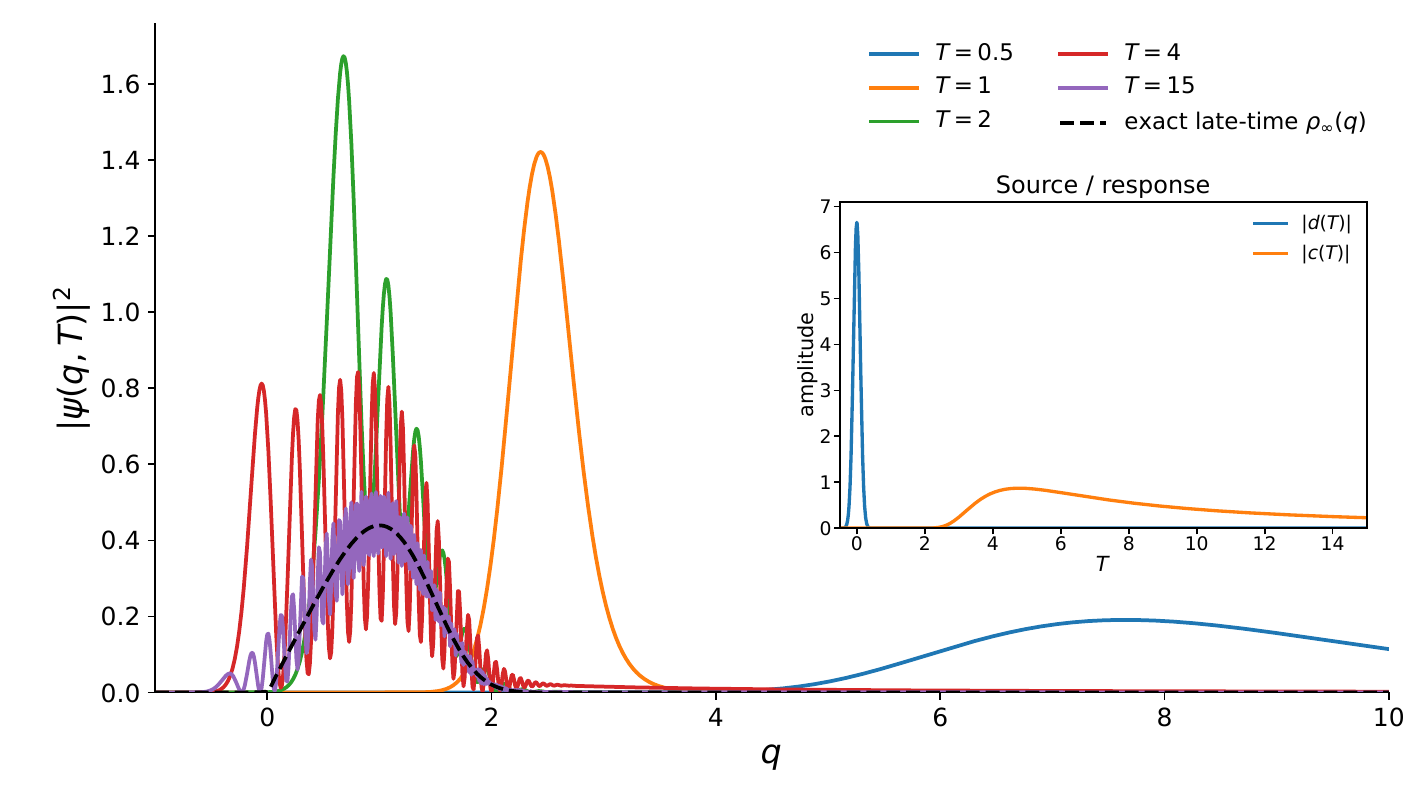}
    \caption{Similar to Fig. \ref{fig:early_time_gaussians_-3} with $T_0 = 0$, $\Delta = 1$, $g=0.1$, and $\epsilon_0 = 0$. This corresponds to a shell with energy distribution centred on the critical threshold. 
    \label{fig:early_time_gaussians_0}
    }
\end{figure}
Now we consider a shell at the critical threshold $\epsilon_0=0$. Fig. \ref{fig:early_time_gaussians_0} shows results for this case. The bounce probability in this case is almost exactly $1/2$ because half of the wavepacket is supported on negative $\epsilon$ (with $|\epsilon| \lesssim \Delta =1$). Classically, a shell with $\epsilon<0$ bounces at $q_b = -1/2-\epsilon$, which probably explains the significant probability density at negative $q$ at intermediate times. The response function shown in the inset peaks around $T =5$, in agreement with the prediction of \eqref{tbounce2}. The wavefunction does not exhibit any pronounced peak corresponding to an outgoing shell; this is presumably because the bounce time diverges as $\epsilon \rightarrow 0-$, which broadens the outgoing wavepacket even more than in Fig. \ref{fig:early_time_gaussians_-3}. 

The Figure exhibits pronounced oscillations which increase in frequency and decrease in amplitude as time increases. These arise from interference between a high-frequency part of the wavefunction describing a shell collapsing to form a black hole and a low frequency part describing a shell that bounces. The frequency of the former increases linearly with $T$ because a shell collapsing to form black hole corresponds to a wavepacket moving outwards with constant velocity at large $Y$, and $|p| \approx Y$ at large $Y$. The amplitude of these oscillations decays as $T^{-3/2}$. This arises from the decay of the (small $p$ hence finite $Y$) ``bounce'' part of the wavefunction, as given by \eqref{psi_late_time}.

As $T\rightarrow \infty$, the profile settles down to the limiting dashed black curve whose form will be determined analytically below. This describes a shell that collapses to form a black hole. The time it takes to ``settle down'' to the limiting profile is much longer than in Fig. \ref{fig:early_time_gaussians_3}. This is presumably because this profile is peaked at $\epsilon=0$ and so the ``late time tail'' of the wavefunction, determined by equation \eqref{psi_late_time}, is much larger in this case. 

Finally we can consider what would happen for more realistic parameter values than shown in the plots. For moderately small $\eta$ we have $g \sim S_0^{-1}$, which is tiny for a macroscopic shell. A Wavepacket localized in time has $\Delta \gtrsim \eta^{-1} g$. If the physical width of the wavepacket is not hugely suppressed by a negative power of $S_0$ then we must have $\Delta \sim \eta^{-1} g$. The negative $\epsilon$ part of the wavepacket then gives a typical bounce time $T_{\rm bounce} \sim 1/\sqrt{\Delta} \sim \sqrt{\eta/g} \sim \eta \sqrt{S_0}$ which is very large (as we saw in \eqref{tbounce}). Hence the interference effect seen in Fig \ref{fig:early_time_gaussians_0} will persist for an incredibly long time for physical parameter values. 

\subsection{Late time behaviour}
\label{sec:late_time_behaviour}

As $T \rightarrow \infty$, any probability associated with a bouncing shell (the response $c(T)$) leaves the system through the AdS$_2$ boundary at $q=\infty$. What remains is the part of the wavefunction describing gravitational collapse to form a black hole. This corresponds to the dashed black curve in Figs \ref{fig:early_time_gaussians_3} and \ref{fig:early_time_gaussians_0}. In this section we will determine this late-time wavefunction for a general source wavepacket $d(T)$. 

Our general solution gives the momentum space wavefunction as
\be
\label{mom_space_wavefn}
 \tilde{\psi}(p,T) = 2\int_{-\infty}^\infty d\epsilon \, \tilde{d}(\epsilon)\exp\left(-\frac{i\epsilon T}{g}-\frac{\pi\sqrt{2\epsilon}}{g}\right)\tilde{\psi}_\epsilon(p)
\ee
where $\tilde{\psi}_\epsilon(p)$ is given in \eqref{eigenstates_p}. As before, we assume that $\tilde{d}(\epsilon)$ is smooth, bounded, and rapidly decaying. One can show that the $\epsilon<0$ contribution to \eqref{mom_space_wavefn} is negligible in the large $T$ limit by an application of the Riemann-Lebesgue lemma (this contribution describes bouncing shells), and we therefore focus on the contribution from $\epsilon>0$. 

At late time, classical trajectories with $\epsilon>0$ have $p\approx \pm \sqrt{2\epsilon}\,T$ where the lower sign corresponds to a shell collapsing to form a black hole and the upper sign to an unphysical trajectory approaching a future inner horizon. Our quantization procedure started by restricting to the physical region of the classical phase space so our physical Hilbert space does not contain unphysical trajectories. However, in order to define $(q,p)$ quantum mechanically we had to enlarge our Hilbert space. The enlarged Hilbert space does contain unphysical trajectories. And even a physical wavefunction, when viewed from the enlarged space, contains a part that ``looks unphysical''. In particular, for any fixed positive energy interval $[\epsilon_-, \epsilon_+]$ with $\epsilon_->0$, the corresponding part of the wavepacket becomes concentrated in the regions $p \lesssim -\sqrt{2\epsilon_-}\,T$ and $p \gtrsim \sqrt{2\epsilon_-}\,T$ at late time. The interpretation of the fact that the wavefunction has support in the latter ``unphysical'' region will be discussed further below. 

This dispersion to large $|p|$ implies that we can obtain asymptotically exact results at late times by focusing on the large-$|p|$ behaviour of the wavefunction. For large $|p|$, we can expand $\tilde{\psi}_\epsilon(p)$ as
\be
\label{approx_eigenfunctions}
 \tilde{\psi}_{\epsilon}(p) \approx \frac{\sqrt{2}}{2}\exp\left[ \left(\frac{i}{g}-\frac{\pi i}{4}(1+\nu)\right)\text{sgn}(p) \right]\frac{g^{-\nu/2}}{\Gamma(1+\nu/2)}e^{-|p|\nu/2}
\ee
and, letting $p=uT$, we can therefore write at large $T$ and fixed $u\neq 0$,
\be
\label{late_time_psi_tilde}
 \tilde{\psi}(uT,T) \approx \int_{0}^{\infty} dk\, k\, C_{\text{sgn}(u)}(k)\exp\left[\frac{iT}{g}\left(k|u|-\frac{k^2}{2}\right)\right]
\ee
where we have set $k = \sqrt{2\epsilon} = ig\nu/2$, and where
\be
 C_{\pm}(k) = \sqrt{2} e^{\pm(i/g - i \pi/4 )}\exp\left(- \frac{\pi k}{g}\left(1\pm\frac{1}{2}\right)\right)\frac{g^{ik/g}}{\Gamma(1-ik/g)}\tilde{d}(k^2/2)
\ee
We now Fourier transform to position space using this large-$|p|$ asymptotic form:
\be
 \psi(q,T) = \frac{T}{\sqrt{2\pi g}}\int_{-\infty}^{\infty} du \, \int_0^{\infty} dk \, k \, C_{\text{sgn}(u)}(k) \exp\left[\frac{iT}{g}\left(k|u| - \frac{k^2}{2} + qu\right)\right] + o(1)
\ee
as $T\to\infty$. Now, using $T/g$ as the large parameter, we apply the stationary phase approximation. The phase is made stationary at
\be
 k_\text{sp} = -q\,\text{sgn}(u) \qquad\qquad u_\text{sp} = -q
\ee
Since $k>0$, this saddle only exists when $\text{sgn}(p) = -\text{sgn}(q)$. We therefore find that
\be
\label{late_time_wavefn}
 \psi(q,T) \approx \sqrt{2\pi g}|q| C_{-\text{sgn}(q)}(|q|)e^{-iq^2T/(2g)}
\ee
It follows that the (unnormalized) probability density function for $q$ in the late-time limit is given by
\be
\label{late_time_distr}
 \rho_{\infty}(q) = 2\pi g \,q^2 |C_{-\text{sgn}(q)}(|q|)|^2
\ee
This simplifies to
\be
\label{rhoinf}
 \rho_{\infty}(q) = 2g^2 \, |q||\tilde{d}(q^2/2)|^2\begin{cases}
     1-e^{-2\pi q/g} & q>0 \\
     e^{-2\pi|q|/g} - e^{-4\pi |q|/g} & q<0
 \end{cases}
\ee
This implies that the probability density for $|q|$ is proportional to 
\be
\label{modqdist}
 \rho_\infty(q) + \rho_\infty(-q) = 2g^2 \, |q||\tilde{d}(q^2/2)|^2 \left( 1-e^{-4\pi |q|/g} \right)
\ee
To interpret this, recall that a black hole of energy $\epsilon$ has inner/outer horizons at $q$ satisfying $\epsilon=q^2/2$. Substituting the latter equation into \eqref{fdef}, the probability density function for the black hole energy, and using $f(\epsilon) d\epsilon = |q| f(q^2/2) d|q|$ reproduces \eqref{modqdist} (up to normalization). Hence \eqref{modqdist} is consistent with the expectation that the location of the shell asymptotes to the location of a black hole horizon as $T \rightarrow \infty$.  

\begin{figure}[htbp]
    \centering
    \includegraphics[width=0.7\textwidth]{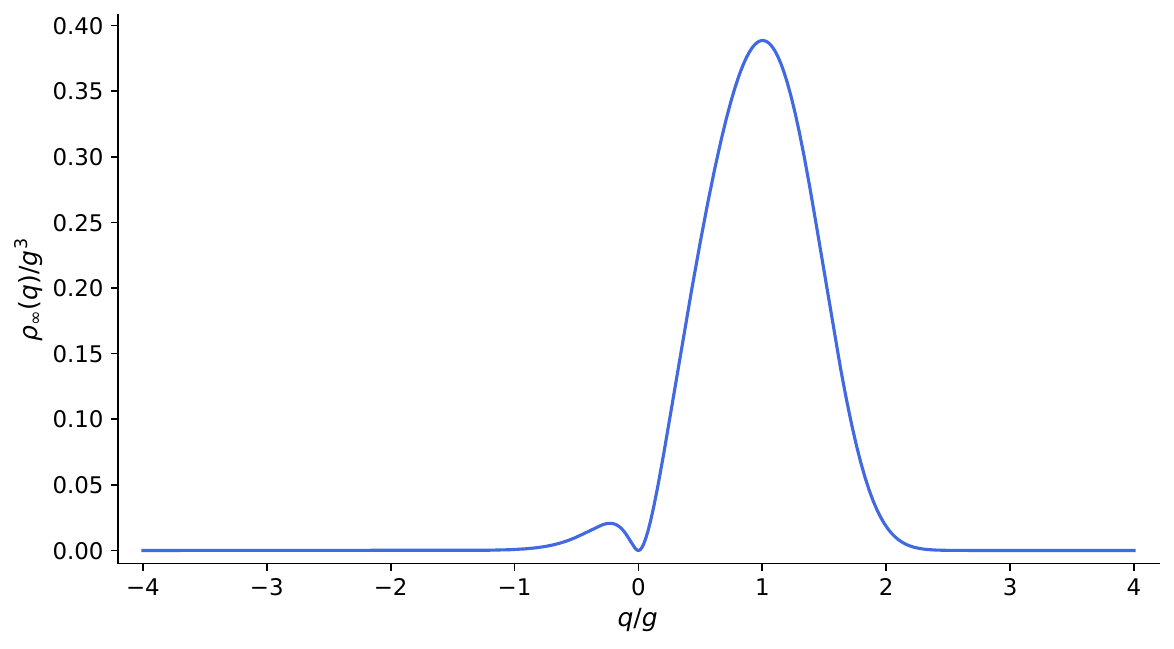}
    \caption{Plot of $\rho_\infty(q)$ for a Gaussian wavepacket with $\Delta/g^2=1$ and $\epsilon_0 = 0$.} 
    \label{fig:rho_infinity}
\end{figure}

The interpretation of \eqref{rhoinf} is less obvious. This gives a non-zero probability density for the shell to be located at {\it negative} $q$ at late time. For example, Fig. \ref{fig:rho_infinity} shows the behaviour of $\rho_\infty(q)$ for a Gaussian wavepacket described in Section \ref{sec:gaussian}, with $\epsilon_0=0$ and $\Delta/g^2=1$. The plot shows a small peak at $q<0$. This seemingly predicts a non-vanishing probability that the shell is located at the unphysical {\it inner} horizon of the black hole. This confirms what we mentioned above when discussing the late time behaviour of the momentum space wavefunction. 

What are we to make of this? We worked hard to exclude the unphysical region from our phase space before quantizing and so it would be disappointing if this region was still hidden in our physical Hilbert space after all! Fortunately, this is not the case as we will now explain.

Let's fix an interval $I = [a,b]$ where $0<a<b$. We will also consider the interval $-I = [-b,-a]$. Consider the probability of measuring $|q| \in I$ at late time. The probability density for this measurement is given in \eqref{modqdist} and, by definition, integrating this over $I$ must agree with $\langle \psi| F_{I\cup(-I)} | \psi \rangle$ because this measurement is equivalent to measuring $q \in I \cup (-I)$. If this measurement produces a positive outcome then the probability that immediately repeating the measurement produces a positive outcome is given by \eqref{repeat} with $I \rightarrow I\cup(-I)$. In Appendix \ref{app:povm} we show that, at late time, this repeat probability is exactly $1$. Hence our measurement prescription produces perfectly repeatable results for measurements of $|q|$. 

Now consider measuring $q$ instead of $|q|$. If we measure whether $q \in \pm I$, obtain a positive outcome, and immediately repeat the measurement then, at late time, the probability of again obtaining a positive outcome is (Appendix \ref{app:povm})
\be
 P_\text{repeat}(q\in\pm I) = \frac{\int_{\pm I} dq\, \frac{\rho_\infty(q)}{1+e^{-2\pi q/g}}}{\int_{\pm I} dq\, \rho_\infty(q)}
\ee
This is always greater than $1/2$ for a measurement corresponding to $I$, and always less than $1/2$ for a measurement corresponding to $-I$. Thus an observation of $q \in -I$ is {\it never} guaranteed to be repeatable. When $a/g\gg 1$, we have $P_\text{repeat}(q \in I) \approx 1$, and $P_\text{repeat}(q\in -I) \approx 0$. So if $a/g \gg 1$ then a measurement of $q \in I$ is almost guaranteed to be repeatable and a measurement of $q \in -I$ is almost guaranteed to be unrepeatable. 

From this we conclude that the possibility of a measurement of negative $q$ does not indicate that the unphysical region is somehow still hidden in our physical Hilbert space. For if it were, then presumably this qualitative statement would be true for any positive energy above extremality. Consider the case of a Gaussian wavepacket with $\epsilon_0 \gg \Delta \gg \epsilon_\star$, describing a shell which forms a black hole with radius $q_+ \approx \sqrt{2 \epsilon_0}$ with high probability. Take $a=0.5\sqrt{2\epsilon_0}$ and $b=1.5\sqrt{2\epsilon_0}$ (say). At late time, the probability of measuring $q \in I$ is extremely close to $1$. But say we measure $q \in -I$. Does this mean the shell has quantum tunneled into the unphysical region of phase space? If it had, then semiclassical reasoning indicates that it would stay in this region. But if we immediately repeat the measurement of $q$ then the probability of again finding $q \in -I$ is miniscule. It is simply that our prescription for measuring the sign of the shell position is inherently a bit fuzzy. 

\section{Discussion}

\label{sec:discuss}

We have considered the spherically symmetric gravitational collapse of a thin shell of charged dust. Classically, the shell will collapse to form a black hole if $E \ge Q$ and bounce without forming a black hole if $E<Q$. If a shell with rest mass close to $Q$ is near the critical threshold then the spacetime geometry outside the shell is AdS$_2 \times S^2$, both above and below the threshold. For such shells, we have shown that the classical motion is described by a Liouville theory with a negative potential. We have quantized this theory as an open system, with ``source'' and ``response'' functions describing the probability current of a shell incident from, and returning to, infinity in AdS$_2$. 

Our theory predicts that a shell whose energy is definitely below $Q$ will definitely bounce without forming a black hole. A shell with $E-Q \gg E_\star$ will form a black hole with high probability. A shell with $0 < E-Q \lesssim E_\star$ might form a black hole, or might undergo a bounce, with the bounce probability approaching $1$ as $E \rightarrow Q+$. Thus the critical surface of the classical theory is thickened quantum mechanically into a ``critical layer'' of thickness $E_\star$. This bounce arises from the collapsing shell quantum tunneling to a trajectory corresponding to a shell emerging from a white hole. 

A shell that is injected at a well-defined time is described by a wavepacket with a spread in energy that is much larger than $E_\star$. If such a wavepacket is centred on the critical threshold $E=Q$ then the probability of black hole formation is $1/2$. One could regard the spread of the wavefunction as producing a perturbation that is equally likely to push the shell above or below the threshold. If a black hole forms then, with high probability, it has $E-Q \gg E_\star$, i.e., it lies outside the critical layer. For a shell of astrophysical size, if a black hole does not form then it takes an incredibly long time for the shell to re-emerge, many times the age of the Universe. The system therefore ``looks like'' a black hole for a very long time. It seems plausible that the features just described would hold also for more realistic matter models such as the Vlasov matter of \cite{Kehle:2024vyt}.

If a black hole does form then our result for the probability distribution of the black hole energy is in excellent agreement with the corresponding result of Turiaci and Wu for collapsing a shell onto a pre-existing near-extremal black hole \cite{Turiaci:2026gtl}. It is reassuring that two different approaches, both based on simplified models, lead to the same result. It would be interesting to understand whether there is some way of connecting these two approaches more directly. The model of \cite{Turiaci:2026gtl} is closely related to ideas in holography. If our model could be related to that of \cite{Turiaci:2026gtl} then perhaps it would give some insight into the holographic description of the flat region inside our shell.

Our model neglects many important physical effects. The restriction to spherical symmetry and a shell of matter implies that no field-theory degrees of freedom are present. Our model does not describe Hawking radiation or Schwinger pair production of charged particles. It would be interesting to see if the model could be extended to include any of these effects.

\subsection*{Acknowledgments}

HSR thanks Gabriele Di Ubaldo, Roberto Emparan, Joaquin Turiaci and the participants of the GGI school ``Pathways to Quantum Black Holes'' for helpful discussions. This work has been partially supported by STFC consolidated grant UKRI4224. HL is supported by a PhD studentship funded by XTX Markets. The authors acknowledge OpenAI for providing access to ChatGPT through the ChatGPT for Academic Researchers programme. ChatGPT was used as a research aid during the development of this work; all results and arguments were independently checked by the authors.

\appendix

\section{Hamiltonian for general shell motion}

\label{app:ham}

In attempting to quantize the motion of the shell, previous authors have tried to express $E$ as a function of the shell's position and velocity, and then pass to a Hamiltonian description so that $E$ is the value of the Hamiltonian. The Hamiltonian should generate time translations but there are several notions of time in this problem: there is the exterior time $t$, there in the interior Minkowski time coordinate and there is the shell's proper time. Different authors have made different choices for which of these should be conjugate to $E$: Ref. \cite{Hajicek:1992mx} uses proper time. Ref. \cite{Hajicek:1992pu} uses the interior Minkowski time. These choices both appear questionable: $E$ is the ADM energy measured at infinity, which suggests that it should generate time translations in $t$. From a more practical perspective, both of these time choices make it difficult to distinguish shell trajectories which form black holes from those that do not: in both cases the shell bounces, in the former case this bounce occurs inside the black hole and the solution extends through the inner horizon into other regions of spacetime. But these two cases do not correspond to any qualitatively different behaviour of the shell's radius as a function of proper time or Minkowski time. 

We are going to assume that $E$ is conjugate to the exterior Reissner-Nordstr\"om time $t$. We aim to find a Hamiltonian $H(r,p_r)$ which generates the equations of motion w.r.t. time $t$ and so that the value of $H$ on-shell is $E$. The shell equation of motion w.r.t. $t$ is
\be
 \left( \frac{dr}{dt} \right)^2 = \frac{\left(1 - \frac{2E}{r} + \frac{Q^2}{r^2} \right)^2 \left[ \left( E-\frac{Q^2-m^2}{2r}\right)^2 - m^2 \right]}{\left(E- \frac{Q^2+m^2}{2r}\right)^2}
\ee
We now replace $dr/dt$ with $\partial H/\partial p_r$ and $E$ with $H$ to obtain
\be
 \frac{\partial H}{\partial p_r} = \pm  \frac{\left(1 - \frac{2H}{r} + \frac{Q^2}{r^2} \right) \left[ \left( H-\frac{Q^2-m^2}{2r}\right)^2 - m^2 \right]^{1/2}}{\left(H- \frac{Q^2+m^2}{2r}\right)}
\ee
Assuming $H > m+(Q^2-m^2)/(2r)$ and setting $H(r,p)=m \cosh \xi(r,p_r)+(Q^2-m^2)/(2r)$ gives
\be
 \pm p_r = \int d\xi \frac{ m \cosh \xi - \frac{m^2}{r}}{1+\frac{m^2}{r^2}-\frac{2m}{r} \cosh \xi}= \frac{r}{2} \log \left| \frac{1-(m/r)e^{-\xi}}{e^\xi -r/m} \right| +f(r)
\ee
where the arbitrary function $f(r)$ can be eliminated by a corresponding shift in $p_r$, which is a canonical transformation, or instead chosen to impose some convenient condition at $r=\infty$. For example, assuming $H$ remains finite as $r \rightarrow \infty$ at fixed $p_r$ requires $f(r) \sim (r/2) \log (r/m)$ as $r \rightarrow \infty$. Making the choice $f(r) = (r/2) \log (r/m)$ leads to
\be
\label{Hfull}
 H = \cosh (p_r/r) \left[r^2 \sinh^2 (p_r/r) + m^2 \right]^{1/2} - r \sinh^2 (p_r/r) +(Q^2-m^2)/(2r)
\ee
To recover the Hamiltonian \eqref{htanh} used in the main body of this paper, substitute \eqref{etadef}, \eqref{qTdef}, $p_r=Q p$ and expand to quadratic order in $\eta$. 

The Hamiltonian \eqref{Hfull} can be simplifed by defining $\nu = r^2/2$ and $\mu = p_r/r$. This gives $d\nu \wedge d\mu = dr\wedge dp_r$ so $(r,p_r) \mapsto (\nu,\mu)$ is a canonical transformation. The Hamiltonian becomes
\be
 H = \cosh \mu \left[2\nu \sinh^2 (\mu) + m^2 \right]^{1/2} - \sqrt{2\nu} \sinh^2 \mu +(Q^2-m^2)/(2\sqrt{2\nu})
\ee
For $Q=0$, this has been derived previously using different methods \cite[eq 43]{kijowski1998true}, \cite[eq 6.4]{Kijowski:2006gd}. 

\section{Origin of the canonical transformation}

\label{app:canonical}

The canonical transformation \eqref{canonical} was discovered using the following sequence of transformations. First the form of the $p$-dependent term in $h$ suggests defining a new momentum
\be
 P = -(q+1) \tanh p 
\ee
and seeking a new position variable $Q$ (not to be confused with the black hole charge) such that $dQ\wedge dP = dq \wedge dp$ leads us to
\be
 Q=  \log \sinh |p|
\ee
This transformation is not globally defined: it is not invertible at $p=0$. So this transformation defines $(Q,P)$ in two different coordinate patches of phase space distinguished by the value of 
\be
 \sigma = {\rm sign}(p)
\ee
In either patch the Hamiltonian becomes
\be  
 h = \frac{P^2}{2} +\sigma P \left( 1 + e^{-2Q} \right)^{1/2} + \frac{1}{2} 
\ee
Completing the square on $P$ suggests performing a further canonical transformation $(Q,P) \rightarrow (Q,\Pi)$ where
\be
 \Pi = P +\sigma  \left( 1 + e^{-2Q} \right)^{1/2}
\ee
and the Hamiltonian becomes
\be
\label{liouville}
 h = \frac{1}{2}\Pi^2 -\frac{1}{2}e^{-2Q}
\ee
which is the Hamiltonian of a particle of unit mass moving in an exponential (Liouville) potential with a negative coefficient. In these coordinates, a bounce has $Q \rightarrow -\infty$, $\sigma \Pi \rightarrow \infty$ with $\sigma \Pi -e^{-Q} \rightarrow 0$. Horizons correspond to $Q \rightarrow \infty$ with $\Pi \rightarrow {\rm constant}$. The physical region of each coordinate patch is given by
\be
 a \equiv  e^{-Q}  -\sigma \Pi>0 \qquad \qquad \Leftrightarrow \qquad \qquad {\rm physical \,\, region} 
\ee
On the physical region we now define
\be
 K = -\sigma a
\ee
so the sign of $K$ determines $\sigma$. On the physical region, defining
\be
 Y = Q-\log a
\ee
gives $dY \wedge dK = dQ \wedge d\Pi$ so the transformation is canonical. Finally reversing the above steps leads back to \eqref{canonical}, which refers to only a single coordinate patch in the original coordinates. 

\section{Rigorous definitions and existence proof}

\subsection{Defining the source and response}

\label{app:cd}

A rigorous definition of $c(T)$ and $d(T)$ can be given as follows. First define the functions
\be
 u(Y) = H_0^{(1)}(z) \qquad \qquad v(Y) = H_0^{(2)}(z)
\ee
where $z$ is defined by \eqref{zdef}. These satisfy
\be
\label{huv}
 h u = h v = 0
\ee
Define the Wronskian of $f(Y)$ and $g(Y)$ as
\be
 \{ f,g\} = f \frac{dg}{dY} - g \frac{df}{dY}
\ee
Equation \eqref{huv} implies that $\{u,v \}$ is constant and evaluating at $Y=-\infty$ gives
\be
\label{wronkuv}
 \{u,v\} = \frac{2i}{\pi}
\ee
For any twice differentiable $\Psi$ we have
\be
 \frac{d}{dY} \{\Psi,v \} = \frac{2}{g^2} (h \Psi) v 
\ee
so integrating gives
\be
 [\{\Psi,v \}]_{Y_0}^{Y_1} = \frac{2}{g^2}\int_{Y_0}^{Y_1} dY  (h \Psi) v 
\ee
Now assume that $h\Psi \in L^2(\mathbb{R})$. $v$ is square integrable at $Y=-\infty$ and so Cauchy-Schwarz implies that the RHS converges as $Y_0 \rightarrow-\infty$. Hence $\lim_{Y \rightarrow -\infty} \{\Psi,v \}$ exists. Similarly $\lim_{Y \rightarrow -\infty} \{u,\Psi \}$ exists. For $\Psi(Y,T)$ that is (almost everywhere) twice-differentiable in $Y$ with $\Psi,h\Psi \in L^2(\mathbb{R})$ we can therefore define
\be
 c(T) =\lim_{Y \rightarrow -\infty} c(Y,T) \qquad \qquad d(T) = \lim_{Y \rightarrow -\infty}d(Y,T)
\ee
where
\be
 c(Y,T) = \frac{\pi}{2i} \{\Psi,v \} \qquad \qquad d(Y,T) = \frac{\pi}{2i} \{u,\Psi \}
\ee
To compute the flux, with the same assumptions about $\Psi$, we first observe that
\be
 \Psi(Y,T) = c(Y,T) u(Y) + d(Y,T) v(Y) \qquad \frac{\partial\Psi}{\partial Y} (Y,T) =  c(Y,T) \frac{du}{dY}(Y) + d(Y,T) \frac{dv}{dY}(Y)
\ee
Substituting this into \eqref{Jdef} and then using $v = \bar{u}$ and \eqref{wronkuv} gives
\be
 J(Y,T) = \frac{g}{\pi} \left( |d(Y,T)|^2 - |c(Y,T)|^2 \right)
\ee
so taking the limit $Y \rightarrow -\infty$ gives \eqref{Jlim}. 

\subsection{Existence proof}

\label{app:existence}

Existence of a solution of the initial-boundary value problem is established as follows. Starting with boundary condition $d(T)=0$, equation \eqref{Jlim} establishes that the time evolution generator $A=-ih/g$ and its adjoint (with boundary condition $c(T)=0$) are both dissipative. A standard contraction semigroup argument then guarantees existence for arbitrary square integrable initial data. With smooth non-zero $d(T)$, subtracting $d(T)w(Y)$ where $w$ equals $H_0^{(2)}$ near $Y=-\infty$ and is smoothly cutoff elsewhere, reduces the problem to the homogeneous boundary condition with a square integrable forcing term, which can be treated using Duhamel's formula. 

\section{Probability of repeat measurements}

\label{app:povm}

Consider the fixed interval $I = [a,b]$ with $0<a<b$, and $-I = [-b,-a]$. We now compute $P_\text{repeat}(q\in\pm I)$ for the late-time physical wavefunctions discussed in Sec. \ref{sec:late_time_behaviour}, as well as the updated wavefunction after a positive outcome for the measurement. We found in \eqref{late_time_wavefn} that the late-time physical wavefunctions under consideration took the form
\be
 \psi(q,T) = A(q)e^{-iq^2T/(2g)}
\ee
up to an error term that vanishes in norm as $T\to\infty$, where $A(q)$ satisfies
\be
\label{A_condition}
 A(-q) = -ie^{2i/g}e^{-\pi q/g}A(q)
\ee
for $q>0$. Let's first consider $F_I\psi(q,T)$ for such wavefunctions at late times. Let $\tilde{\psi}(p,T)$ be the Fourier transform of $\psi(q,T)$, and let $\phi_I(p,T) = R_I\tilde{\psi}(p,T)$. Throughout the rest of this section, we write $\approx$ to mean equality up to an error that vanishes in norm as $T\to\infty$. We have
\be
 \phi_I(p,T) \approx \frac{1}{\sqrt{2\pi g}}\int_a^b dq\, A(q)\exp\left(-\frac{i}{g}\left(pq+\frac{q^2T}{2}\right)\right)
\ee
One can show, using an integration by parts, that
\be
 \int_{-|a|T/2}^{\infty} dp \, |\phi_I(p,T)|^2 = O(g/T)
\ee
as $T\to \infty$ at fixed $b$. We want to compute
\be
 (PR_IP\tilde{\psi})(p,T) = (P\phi_I)(p,T) = \int_{-\infty}^{\infty}dp'\, P(p,p')\phi_I(p',T)
\ee
and it follows that we can get asymptotically exact results by expanding the projection kernel $P(p,p')$ for large negative $p'$. One can show that, as $T\to\infty$,
\be
 \int_{-|a|T}^{|a|T}dp\,|(P\phi_I)(p,T)|^2 \to 0
\ee
and therefore, at large $T$, the only region in which $(P\phi_I)(p,T)$ has non-vanishing norm is $|p|>|a|T$. In the case that $p<-|a|T$ for large $T$, we can expand the projection kernel to find that
\be
 (P\phi_I)(p,T) \approx \frac{1}{2}\phi_I(p,T) - \frac{i}{4\pi}\text{PV}\int_{-\infty}^{\infty}dp'\, \frac{\phi_I(p',T)}{\sinh\left(\frac{p-p'}{2}\right)}
\ee
where the principal value is taken at $p=p'$. Similarly, for $p>|a|T$ for large $T$, we find that
\be
 (P\phi_I)(p,T) \approx -\frac{i}{4\pi}e^{2i/g}\int_{-\infty}^{\infty}dp'\, \frac{\phi_I(p',T)}{\cosh\left(\frac{p+p'}{2}\right)}
\ee
Calling these large negative and positive $p$ approximations $G_-$ and $G_+$ respectively, we can then write
\be
 (P\phi_I)(p,T) \approx \textbf{1}_{p<-|a|T}\,G_-(p,T) + \textbf{1}_{p>|a|T}\,G_+(p,T)
\ee
One can show that, as $T\to\infty$, 
\be
 \int_{-\infty}^{|a|T}dp\,|G_+(p,T)|^2 \to 0
\ee
and the analogous result holds for $G_-$. This allows us to remove the above indicator functions, and Fourier transform $(P\phi_I)(p,T)$ back to position space to find that
\be
 (F_I\psi)(q,T) \approx \textbf{1}_{I}(q)\frac{1}{1+e^{-2\pi q/g}}\psi(q,T) - \textbf{1}_{-I}(q)\frac{ie^{2i/g}}{2\cosh(\pi q/g)}\psi(-q,T)
\ee
as $T\to\infty$. Using the condition \eqref{A_condition}, we find that this reduces to
\be
 (F_I\psi)(q,T) \approx \frac{\textbf{1}_{I\cup (-I)}(q)}{1+e^{-2\pi |q|/g}}\psi(q,T)
\ee
A very similar calculation shows that
\be
 (F_{-I}\psi)(q,T) \approx \frac{\textbf{1}_{I\cup (-I)}(q)}{1+e^{2\pi |q|/g}}\psi(q,T)
\ee
when the condition \eqref{A_condition} holds. Now, notice that we have
\be
 (F_{\pm I}\psi)(q,T) = f_{\pm I}(q)A(q)e^{-iq^2T/(2g)}
\ee
for an even function $f_{\pm I}(q)$ in the late-time limit. It follows that $f_{\pm I}(q)A(q)$ satisfies the condition \eqref{A_condition}, and hence we can see inductively that
\be
 (F_{\pm I}^n\psi)(q,T) = f_{\pm I}(q)^n A(q)e^{-iq^2 T/(2g)}
\ee
for $n\in\mathbb{N}$ in the late-time limit. It is then straightforward to compute
\be
 \langle F_{\pm I}^2 \rangle = \langle \psi | PR_{\pm I}PR_{\pm I}P\psi\rangle = \langle R_{\pm I}\psi | F_{\pm I} \psi \rangle = \int_{\pm I} dq \, \frac{|\psi(q,T)|^2}{1+e^{-2\pi q/g}} + o(1)
\ee
as $T\to\infty$, where we use the fact that $P|\psi\rangle = |\psi\rangle$ for physical states. This gives 
\be
 P_\text{repeat}(q\in\pm I) = \frac{\int_{\pm I} dq\, \frac{\rho_\infty(q)}{1+e^{-2\pi q/g}}}{\int_{\pm I} dq\, \rho_\infty(q)}
\ee
We therefore see that $P_\text{repeat}$ is always greater than $1/2$ for a measurement corresponding to $I$, and always less than $1/2$ for a measurement corresponding to $-I$. When $a/g\gg 1$, we have $P_\text{repeat}(q \in I) \approx 1$, and $P_\text{repeat}(q\in -I) \approx 0$. This means that a measurement of $q\in I$ is almost guaranteed to be repeatable in this case, and a measurement of $q \in -I$ is extremely unlikely to be repeatable.

We can also compute $P_\text{repeat}(|q|\in I)$. This follows from the fact that
\be
 (F_{I \cup (-I)}\psi)(q,T) = \left((F_I + F_{-I})\psi\right)(q,T) \approx \textbf{1}_{I \cup (-I)}(q)\psi(q,T)
\ee
at late times, and therefore 
\be
 \langle F_{I\cup (-I)}^2 \rangle = \langle R_{I \cup (\pm I)}\psi | F_{I \cup (\pm I)} \psi \rangle = \int_{I\cup(-I)} dq \, |\psi(q,T)|^2 + o(1)
\ee
as $T\to\infty$, and so 
\be
 P_\text{repeat}(|q| \in I) = 1
\ee
at late times. This shows that, at late times, the measurement process gives perfectly reliable results for $|q|$.

We can also compute $\sqrt{F_{\pm I}}\psi(q,T)$ using the above, in order to determine the updated wavefunction after a positive outcome is achieved for a measurement. This follows by writing 
\be
 \sqrt{F_{\pm I}}\psi(q,T) = (\mathbb{I}-(\mathbb{I}-F_{\pm I}))^{1/2}\psi(q,T)
\ee
and expanding in $(\mathbb{I}-F_{\pm I})$, using the fact that the spectrum of $F_{\pm I}$ lies in $[0,1]$. It is then easy to see that
\be
 \sqrt{F_{\pm I}}\,\psi(q,T) = \sqrt{f_{\pm I}}\,\psi(q,T) + o(1)
\ee
as $T\to \infty$. After normalisation, this gives the form of the updated wavefunction after a measurement of $q \in \pm I$ yields a positive outcome.

\bibliographystyle{utphys}
\bibliography{biblio.bib}

\end{document}